\documentclass[10pt, conference, letterpaper]{IEEEtran}

\usepackage{cite}
\usepackage{subfigure}
\usepackage{url}
\usepackage{verbatim,epsf}
\usepackage[font=footnotesize]{subfig}
\usepackage{enumitem}

\usepackage{multicol}
\usepackage{multirow}
\usepackage[font=footnotesize,labelfont=bf]{
	caption}
\def\tablename{Table}
\usepackage{siunitx}
\usepackage{bm}
\usepackage{tabularx}
\usepackage{balance} %MAU this is needed to balance the column of the last page

\usepackage{amssymb}
\usepackage{amsmath}
\usepackage{graphicx}
\usepackage{epstopdf}

\usepackage{epsfig, color}

\usepackage{comment} % Package per i commenti
\usepackage{amsfonts} % Package per i caratteri  matematici spciali
\usepackage{amsmath} % Package per i caratteri  matematici spciali
\usepackage{rotating}

\usepackage{amssymb}
\usepackage{amsmath}
\usepackage{amsfonts}
\usepackage{wasysym}

\usepackage{algorithm}
\usepackage{algorithmic}

\usepackage{color}

\newsavebox{\ieeealgbox}

\newlength\myindent
\IEEEaftertitletext{\vspace{-0.28in}}

\begin{document}	
\IEEEoverridecommandlockouts

\title{\huge
	{3DRIMR: 3D Reconstruction and Imaging via mmWave Radar based on Deep Learning}}

\author{
	\IEEEauthorblockN{Yue Sun\IEEEauthorrefmark{2}, Zhuoming Huang\IEEEauthorrefmark{1}, 
		Honggang Zhang\IEEEauthorrefmark{1}, Zhi Cao\IEEEauthorrefmark{2}, Deqiang Xu\IEEEauthorrefmark{1}\\
		\IEEEauthorrefmark{1}Engineering Dept., UMass Boston. 
		\IEEEauthorrefmark{2}CS Dept., UMass Boston.\\
		{\{yue.sun001, zhuoming.huang001, honggang.zhang, zhi.cao001, deqiang.xu001\}@umb.edu}}
}

\maketitle
%Add space between copyright and text
\IEEEpubidadjcol

\pagestyle{plain}

\begin{abstract}

%Recent research has shown that 
mmWave radar has been shown as an effective sensing technique 
in low visibility, smoke, dusty, and dense fog environment.
However tapping the potential of radar sensing to reconstruct 3D object shapes remains
a great challenge, due to the characteristics of radar data such as 
sparsity, low resolution, specularity, high noise, and multi-path induced shadow reflections and artifacts. In this paper we propose \textbf{3D} \textbf{R}econstruction and \textbf{I}maging via 
\textbf{m}mWave \textbf{R}adar (3DRIMR), a deep learning based architecture that reconstructs 3D shape of 
an object in dense detailed point cloud format, based on sparse raw mmWave radar intensity data. 
The architecture consists of two back-to-back conditional GAN deep neural networks:
the first generator network generates 2D depth images based on raw radar intensity data, 
and the second generator network outputs 3D point clouds based on the results of the first generator.  
The architecture exploits both convolutional neural network's convolutional operation 
(that extracts local structure neighborhood information)
and the efficiency and detailed geometry capture capability of point clouds (other than
costly voxelization of 3D space or distance fields). 
Our experiments have demonstrated 3DRIMR's effectiveness in reconstructing 3D objects, and 
its performance improvement over standard techniques.

%Keywords: Edge Computing, Edge-Cloud System, Shapley Value, Nash Equilibrium

\end{abstract}

\section{Introduction}

Recent advances in Millimeter Wave (mmWave) radar sensing technology have made it a great tool in autonomous 
vehicles \cite{HawkEye} and search/rescue in high risk areas \cite{mobisys20smoke}. For example, 
in the areas of fire with smoke and toxic gas and hence low visibility,
it is impossible to utilize optical sensors such as LiDAR and camera 
to find a safe route, and any time delay can potentially cost lives in those rescue scenes. 
To construct maps in those heavy smoke and dense fog environments,
mmWave radar has been shown as an effective sensing tool \cite{mobisys20smoke,HawkEye}.

However it is challenging to use mmWave radar signals for object imaging and reconstruction, as
their signals are usually of low resolution, sparse, and highly noisy due to multi-path and 
specularity. A few recent work \cite{HawkEye, superrf,mobisys20smoke} have made some progress
in generating 2D images based on mmWave radar. In this paper, we move
one step further to tackle a more challenging problem: reconstructing 3D object shapes based on raw sparse and 
low-resolution mmWave radar signals. 

To address the challenge, we propose 3DRIMR, a deep neural network architecture (based on conditional GAN)
that takes as input raw mmWave radar sensing signals scanned from multiple 
different viewpoints of an object,
and it outputs the 3D shape of the object in the format of point cloud.  
The architecture consists of two generator networks in two stages. 
In Stage 1, generator network $\bf{G_{r2i}}$ takes 3D radar intensity data as inputs and generates 2D depth images.
In Stage 2, generator network $\bf{G_{p2p}}$ takes as input a set of four 2D depth images 
(results from Stage 1) of the object (from four different viewpoints) 
and generates a dense smooth 3D point cloud of the object.
Generator $\bf{G_{r2i}}$ is jointly trained with 
a discriminator network $\bf{D_{r2i}}$, which fuses 3D radar intensity data and 2D depth images (either
generated or ground truth images).
In addition, generator $\bf{G_{p2p}}$ is jointly trained with 
a discriminator network $\bf{D_{p2p}}$.

3DRIMR architecture is designed to combine the advantages of Convolutional Neural Network (CNN)'s 
convolutional operation
and the efficiency of point cloud representation of 3D objects.  
Convolutional operation can capture detailed local neighborhood structure of a 3D object, 
and point cloud format of 3D objects is more efficient and of higher resolution than 3D
shape representation via voxelization. 
Specifically, $\bf{G_{r2i}}$ applies 3D convolution operation to 3D radar intensity data 
to generate depth images. 
Generator $\bf{G_{p2p}}$ represents a 3D object in point cloud format (unordered point set hence convolutional operation not applicable), so that it can 
process data highly efficiently (in terms of computation and memory) and can express
fine details of 3D geometry.

In addition, because commodity mmWave radar sensors (e.g., TI's IWR6843ISK \cite{iwr6843}) 
usually have good resolution along range direction even without time consuming Synthetic Aperture Radar (SAR) operation, a 2D depth image generated by $\bf{G_{r2i}}$ 
can give us high resolution depth information from the viewpoint
of a radar sensor. Therefore, we believe that combining multiple such 2D depth images from multiple viewpoints
can give us high resolution 3D shape information of an object. 
Our architecture design takes advantage of this observation by using
multiple 2D depth images of an object from multiple viewpoints 
to form a raw point cloud and uses that as input to generator $\bf{G_{p2p}}$.

Our major contributions are as follows: 
\begin{enumerate}
	\item 
	A novel conditional GAN architecture that exploits 3D CNN for local structural information extraction
	and point cloud based neural network for highly efficient 3D shape generation with detailed geometry.  
	
	\item 
	A fast 3D object reconstruction system that implements our architecture and   
	it works on the signals of two radar snapshots of an object by a commodity 
	mmWave radar sensor, instead of a slow full-scale SAR scan.

	\item Our system works directly on sparse and noisy raw radar data without any structural assumption or annotation. 
\end{enumerate}

In the rest of the paper, we briefly discuss related work and background of this work in Sections 
\ref{sec_related} and \ref{sec_background}.
Then we present the design of 3DRIMR in Section \ref{sec_design}. 
Implementation details and experiment results are given in 
Section \ref{sec_imp}. 
Finally, the paper concludes in Section \ref{sec_conclusion}.

\section{Related Work}\label{sec_related}

%\noindent \textbf{mmWave Radar sensing and imaging.} 
There have been active research in recent years on 
applying Frequency Modulated Continuous Wave (FMCW) Millimeter Wave (mmWave) radar sensing in 
many application scenarios, for example, person/gesture identification \cite{vandersmissen2018indoor,yang2020mu}, 
car detection/imaging \cite{HawkEye} and environment sensing \cite{mobisys20smoke, superrf} in low visibility environment. 
To achieve high resolution radar imaging, usually mmWave radar systems need to 
work at close distance to objects or they rely on SAR process \cite{mamandipoor201460,national2018airport,ghasr2016wideband,sheen2007near}.

Our work is inspired by a few recent researches on mmWave radar imaging, mapping, and 3D object 
reconstruction \cite{HawkEye,superrf,mobisys20smoke,yuan2018pcn,qi2016pointnet}. 
Both \cite{HawkEye,mobisys20smoke} have shown the capability of 
mmWave radar in sensing/imaging in low visibility environment.
%, e.g., penetrating dense fog or smoke.
Their findings motivate us to adopt mmWave radar in our design. In addition, we intend to 
add the low-visibility sensing capability on our UAV SLAM system \cite{sun2020lidaus} for search and rescue 
in dangerous environment.
% thus mmWave radar is our natural choice. 

%\noindent \textbf{3D Object Reconstruction.} 
Learning-based approaches have been adopted in recent research on 3D object shape reconstruction \cite{yang20173d,dai2017shape,sharma2016vconv,smith2017improved}.
Most of them use voxels to represent 3D objects, 
as CNN convolutional operation can be easily applied to such data format. However, voxelization of 
3D objects or space cannot achieve high resolution due to cubic growth of memory and computation cost.
Our architecture's generator at Stage 2 
uses point cloud to represent 3D objects which can give us detailed geometric information 
with efficient memory and computation performance. In addition, in Stage 1 of our architecture, 
we are able to take advantage of convolutional operation to directly work on 3D radar intensity data.

HawkEye \cite{HawkEye} generates 2D depth images of an object based on 
conditional GAN architecture with 3D radar intensity maps obtained by multiple SAR scans along both elevation and azimuth dimensions. Our architecture adopts this design to generate intermediate results used as inputs to a 3D point cloud generator. 
Different from \cite{HawkEye}, we use data of only two snapshots 
%from only one radar scan 
along elevation dimension
% which gives  from two virtual antenna receivers  
when using commodity TI IWR6843ISK sensor \cite{iwr6843}. 
This is similar to the input data used in \cite{superrf}, but the network of \cite{superrf}
outputs just a higher dimensional radar intensity map along elevation dimension, not 2D depth images. 

The research in \cite{yuan2018pcn} and \cite{qi2016pointnet} uses point cloud to reconstruct 3D object shapes, 
which motivates us to adopt PointNet structure in our 3D point cloud generator. Different from those existing work, our architecture's Stage 2 utilizes a conditional GAN architecture to jointly train a generator and a discriminator to 
achieve high prediction accuracy. 

\section{Background}\label{sec_background}

\subsection{FMCW Millimeter Wave Radar Sensing and Imaging}
Frequency Modulated Continuous Wave (FMCW) mmWave radar sensor works 
by periodically transmitting continuous chirps that each linearly sweeps through 
a certain frequency band \cite{timmwave}. 
Transmitted signals reflected back from an object in a 3D space will be received 
by a receiver. Range Fast Fourier Transform (FFT) is conducted on the received waveforms 
to detect the distance of the object from the radar sensor. 
In addition, multiple receivers can be arranged horizontally and vertically
%  used along elevation and azimuth dimension, and their received signals can 
to form virtual antenna arrays along both dimensions. Then  
two additional FFTs can be applied to the data to calculate an object's relative angles from 
the sensor horizontally and vertically, referred to as azimuth angle $\phi$ 
and elevation angle $\theta$.   
Those three FFTs together can generate a 3D heatmap or intensity map of the space that 
represents the energy or radar intensity per voxel, which is written as $x(\phi, \theta, \rho)$.
%, denoting
%azimuth angle, elevation angle, and distance measured from the sensor.  

The process of electronically or mechanically steering an antenna array
to get high azimuth and elevation angle resolutions 
%$\delta \phi$ 
%and elevation angle resolution 
%$\delta \theta$ 
is referred to as Synthetic Aperture Radar (SAR) operation. 
Usually higher resolutions along these two dimensions requires longer SAR operation time given fixed number 
of transmitters and receivers. However, high range resolution can be achieved even 
with commodity mmWave radar sensors
(e.g., IWR6843ISK \cite{iwr6843}) without time consuming SAR process. 
In our work, we use IWR6843ISK \cite{iwr6843} operating at $60$ GHz frequency. 

Different from LiDAR and camera sensors, the data generated by mmWave radar sensors is usually 
sparse, of low resolution, and highly noisy.
Even though SAR can help improve resolution, it is a very slow process, 
which may not be practical in many application scenarios that require short application response time. 
The specularity characteristic makes an object's surface behave like a mirror, so the reflected signals from 
a certain portion of the object will not be received by a receiver (hence missing data). 
In addition, the multi-path effect can cause ghost points which give incorrect information on
the object's shape. 
For detailed discussions on FMCW mmWave radar sensing, please see \cite{HawkEye,superrf,mobisys20smoke}.

%\subsection{SAR Imaging}

\subsection{3D Reconstruction}

In this work, we use point cloud, a widely used format in robotics, to represent 3D objects' geometry (e.g., \cite{mobisys20smoke}). This format has been used in recent work 
on learning-based 3D reconstruction, e.g., \cite{fan2017point,qi2016pointnet,qi2017pointnet++}.
A point cloud representation of an object is a set of unordered points, with each point is a sample point of 
the object's surface, written as its 3D Cartesian coordinates.  
Unlike voxelization of 3D objects, a point cloud can represent 
an object with high resolution but without high memory cost. 
However, CNN convolutional operation cannot be applied to an unordered point cloud set. 
% as there is no 
%definition of local neighborhood in a point cloud set. 

In addition, even though we can generate the point cloud of an object by directly filtering out 
mmWave radar 3D heatmaps, such a resulting point cloud usually has very low resolution, being sparse, 
and with incorrect ghost points due to multi-path effect. Therefore even though it may be acceptable to
just use such a point cloud to detect 
the presence of an object, it is impossible 
to reconstruct the shape of the object. Our work attempts to solve 
this problem by using two generator neural networks  
to produce a smooth dense point cloud based on raw radar data. 

Other than point clouds, voxel representations are commonly used in 3D reconstruction to represent 3D objects in learning based approaches, 
for example, \cite{kar2017learning,paschalidou2018raynet,ji2017surfacenet,wu2016learning}, and  
3D CNN convolutional operations can be applied to such data models.
% in a similar way to 
%2D CNN for 2D images. 
However such representations have cubic growth rate in the number of voxel grids, so they 
are limited to representing low resolution objects.
%Some existing research also considers 3D reconstruction based on data captured multiple viewpoints 
%\cite{paschalidou2018raynet, ji2017surfacenet,kar2017learning}. 
In addition, mesh representations of 3D objects are considered in existing work, e.g., 
\cite{kong2017using,wang2018pixel2mesh}, but they are also limited by memory cost and are prone to self-intersecting meshes. 
%\cite{mescheder2019occupancy}

\section{3DRIMR Architecture}\label{sec_design}

\subsection{Overview}

\begin{figure*}[htb!]
	\centerline{
		\begin{minipage}{3.5in}
			\begin{center}
				\setlength{\epsfxsize}{3.4in}
				\epsffile{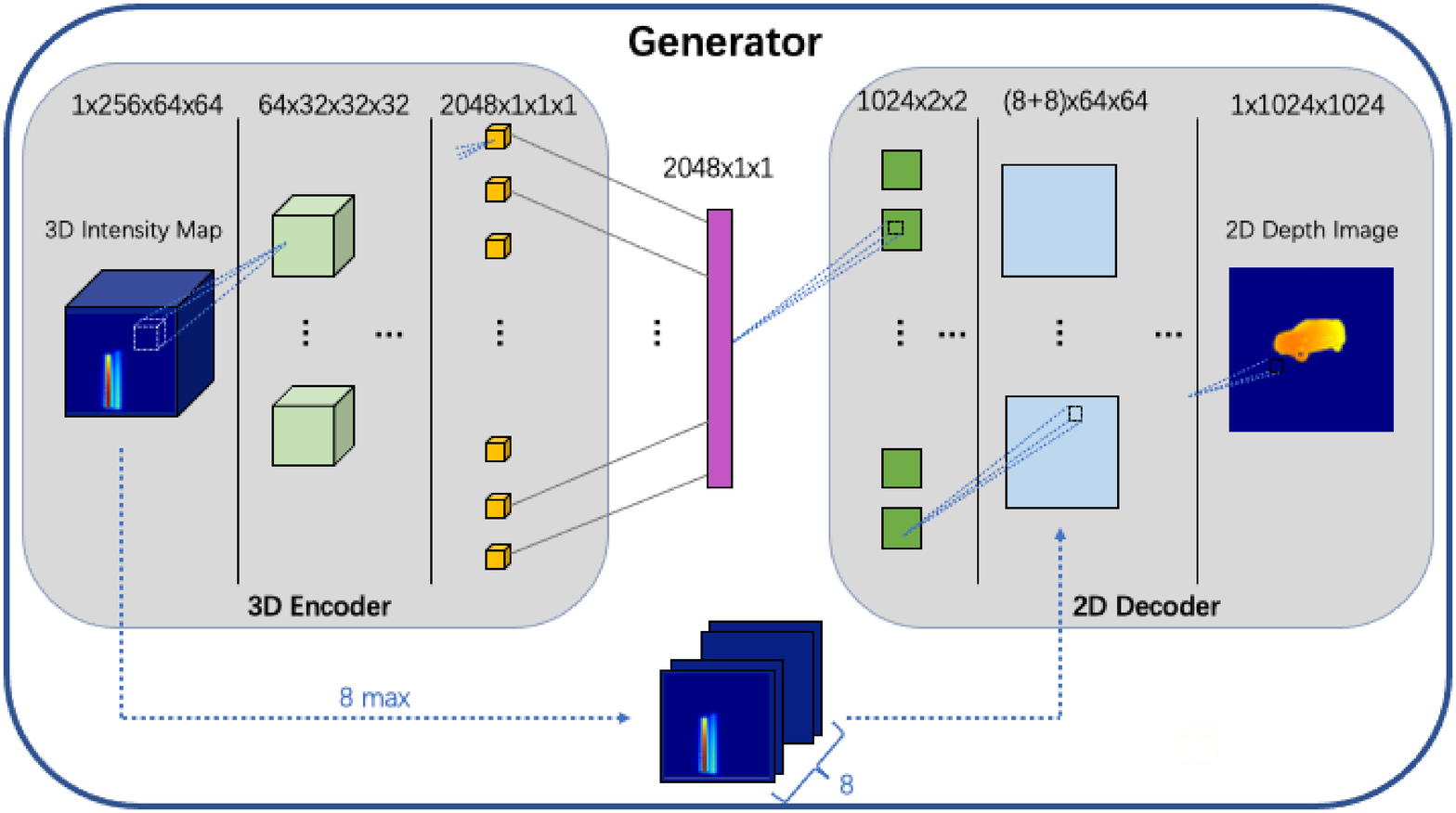}\\
				{}
			\end{center}
		\end{minipage}
			\begin{minipage}{3.5in}
				\begin{center}
					\setlength{\epsfxsize}{3.4in}
					\epsffile{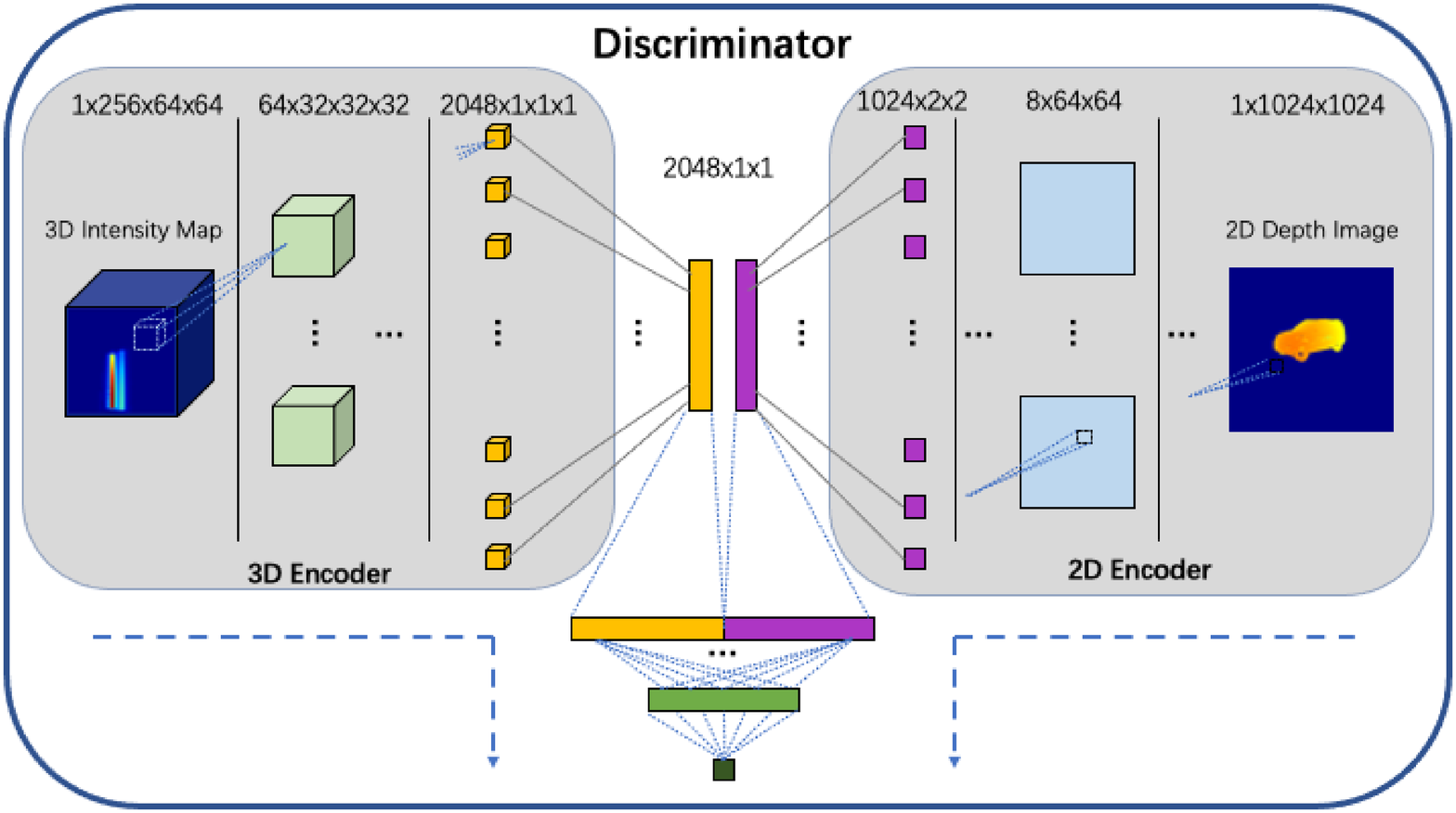}\\
					{}
				\end{center}
			\end{minipage}
	}
	\caption{3DRIMR Stage 1. The generator $\mathbf{G_{r2i}}$ takes an object's 3D radar intensity map as input and generates a 2D depth image. The discriminator $\mathbf{D_{r2i}}$'s input includes 
		a 3D radar intensity map and
		a ground truth depth image or a 
generated 2D depth image. A conditional GAN architecture is used to train both generator and discriminator jointly.}
	\label{fig_stage_1}
\end{figure*}

\begin{figure*}[htb!]
	\centerline{
		\begin{minipage}{3.5in}
			\begin{center}
				\setlength{\epsfxsize}{3.4in}
				\epsffile{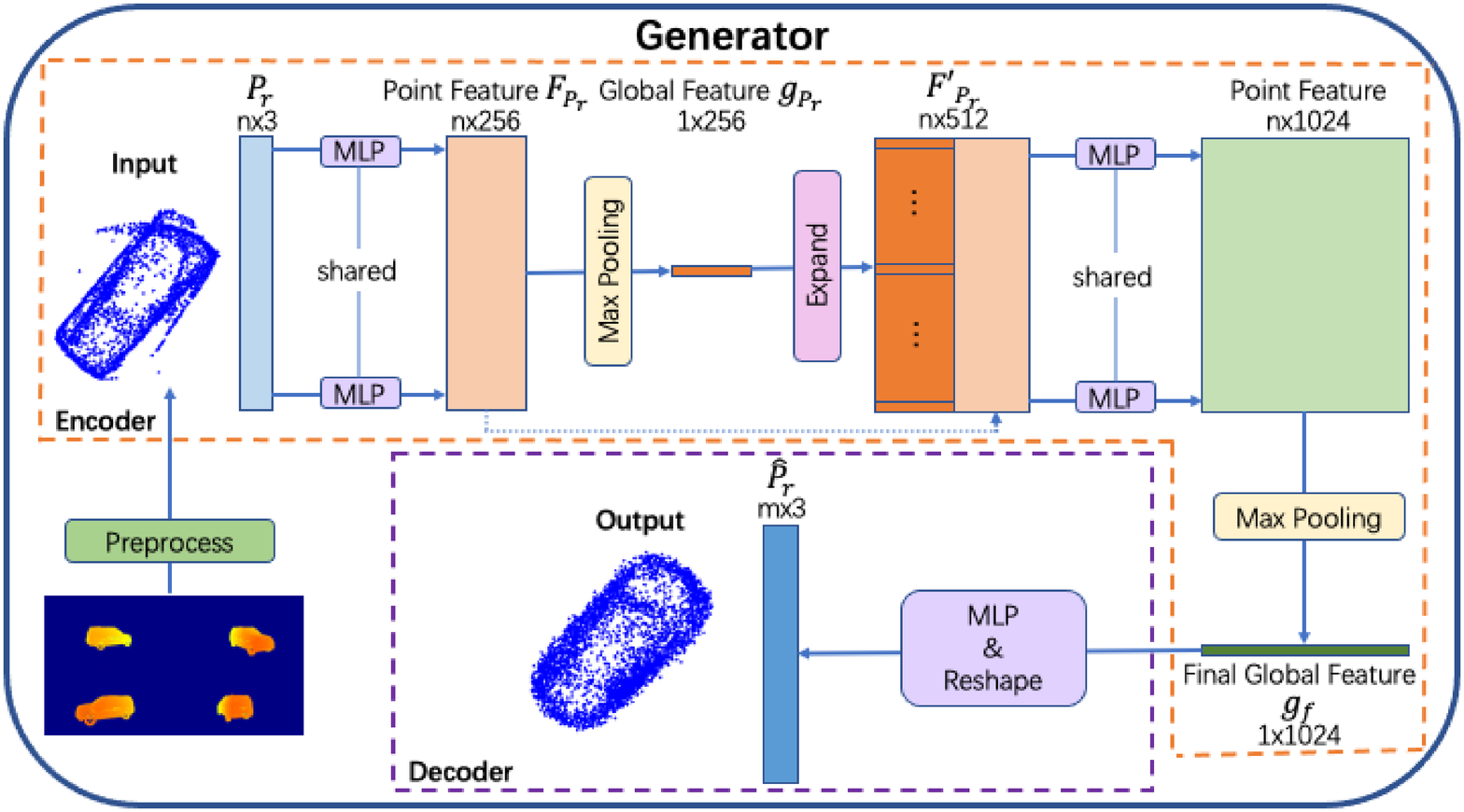}\\
				{}
			\end{center}
		\end{minipage}
		\begin{minipage}{3.5in}
			\begin{center}
				\setlength{\epsfxsize}{3.4in}
				\epsffile{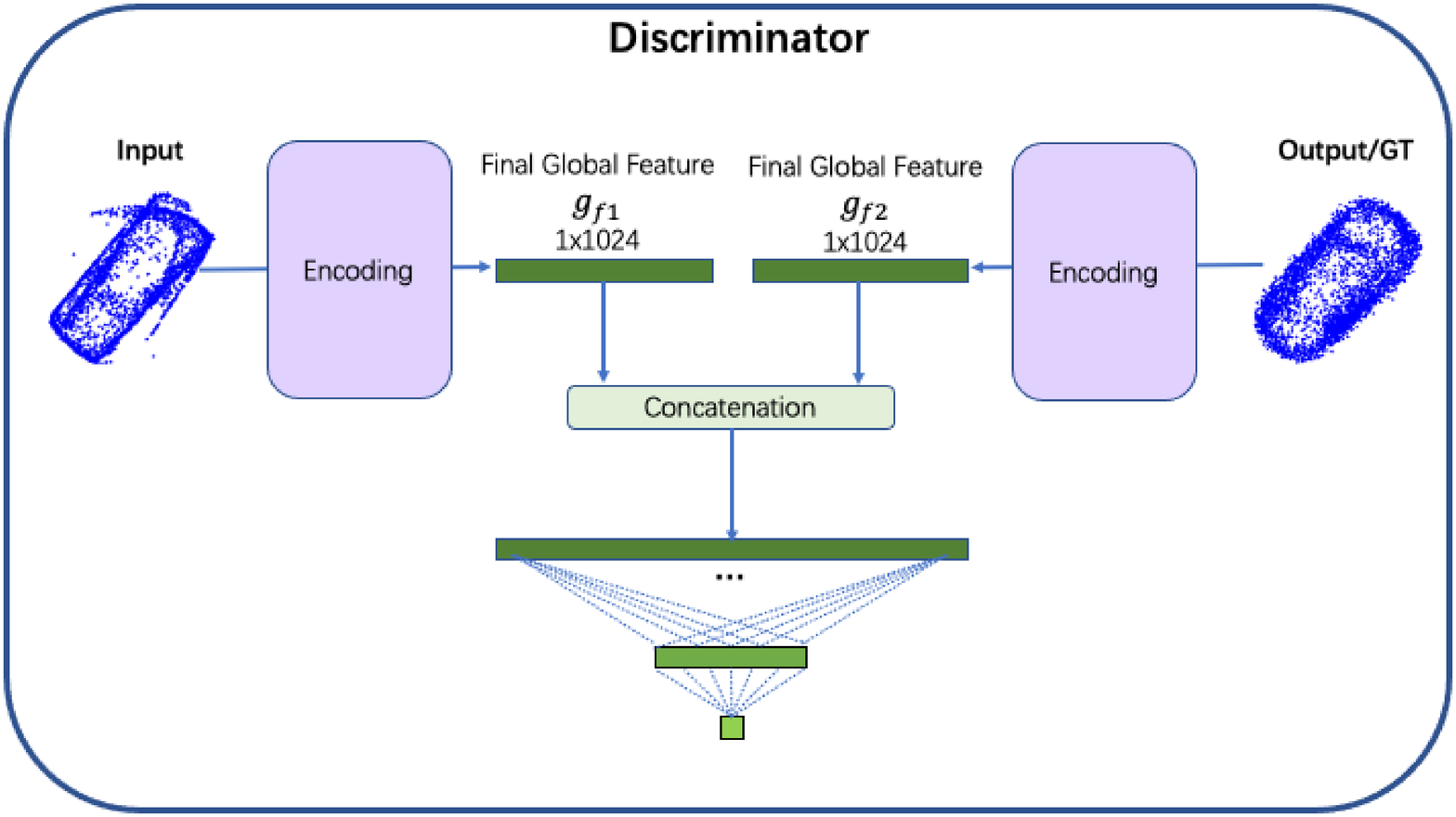}\\
				{}
			\end{center}
		\end{minipage}
	}
	\caption{3DRIMR Stage 2. The system first pre-processes four generated 2D depth images of an object (from Stage 1)   to generate four coarse point clouds. 
	Those four depth images are generated for the same object but viewed from four different viewpoints.
	Combining those four coarse point clouds we can derive a single coarse point cloud, 
	which is used as the input to generator $\mathbf{G_{p2p}}$ in this stage.
	The generator outputs a dense and smooth point cloud representation of the object. 
	A conditional GAN architecture is used to train both generator $\mathbf{G_{p2p}}$ 
	and discriminator $\mathbf{D_{p2p}}$ jointly.}
	\label{fig_stage_2}
\end{figure*}

3DRIMR is a conditional GAN based architecture that consists of two generator networks 
$\mathbf{G_{r2i}}$ and $\mathbf{G_{p2p}}$, 
and two discriminator networks $\mathbf{D_{r2i}}$ and 
$\mathbf{D_{p2p}}$ that are jointly trained together with their corresponding generator networks, 
as shown in Figures \ref{fig_stage_1} and \ref{fig_stage_2}.

Generator $\mathbf{G_{r2i}}$ takes as input a 3D radar energy intensity map
of an object and generates a 2D depth image of the object.
Let $m_r$ denote a 3D radar intensity map of an object captured 
%at position $s$ 
from viewpoint $v$.
Let $g_{2d}$ be a ground truth 2D depth image of the same object captured 
%at the same position and 
from the same viewpoint. 
$\mathbf{G_{r2i}}$ generates $\hat{g}_{2d}$ that predicts or estimates $g_{2d}$ given $m_{r}$. 

Given a set of 3D radar intensity maps $\{m_{r,i} | i = 1, ..., k\}$ of an object captured from $k$ different viewpoints $v_1, ..., v_k$,
generator $\mathbf{G_{r2i}}$ predicts their corresponding 2D depth images $\{\hat{g}_{2d,i} | i = 1, ..., k\}$. 
Each predicted image $\hat{g}_{2d,i}$ can be projected into 3D space to generate a 3D point cloud. 
%with its position and camera setting. 

Our system (that implements 3DRIMR) generates a set of 
$k$ coarse point clouds $\{P_{r,i} | i = 1, ..., k\}$
of the object from $k$ viewpoints. 
%This is a preprocess step needed by generator $\mathbf{G_2}$
%to get its input. 
The system unions the $k$ coarse point clouds 
%from $k$ viewpoints of an object 
to form an initial estimated coarse point cloud of the object, denoted as $P_r$, which is 
a set of 3D points $\{p_j | j = 1, ..., n\}$, and each point $p_j$ 
is a vector of Cartesian coordinates.
% $(x, y, z)$.
We choose $k=4$ in our experiments.
Generator $\mathbf{G_{p2p}}$ takes 
%the initial estimated point cloud 
$P_r$ as input, 
and predicts a dense, smooth, and accurate point cloud $\hat{P}_r$. 

Since the prediction of $\mathbf{G_{r2i}}$ may not be completely correct, a coarse 
$P_r$ may likely contain many missing or even incorrect points. 
The capability of our 3DRIMR handling incorrect information differs our work from 
%This problem has not been addressed 
%in existing work. For instance, 
\cite{yuan2018pcn} which only considers the case of missing points in 
an input point cloud.

By leveraging conditional GAN,
the architecture's training process consists of two stages:
$\mathbf{G_{r2i}}$ and $\mathbf{D_{r2i}}$ are trained together
in Stage 1; similarly, $\mathbf{G_{p2p}}$ and $\mathbf{D_{p2p}}$ are trained together
in Stage 2. The network architecture and training process 
are described in more details in Section \ref{sec_stage1}.

The design of $\mathbf{G_{r2i}}$ is similar to HawkEye \cite{HawkEye}, 
but $\mathbf{G_{r2i}}$'s each input radar intensity map only contains two snapshots
%) along elevation dimension, 
whereas HawkEye's input radar intensity map has $64$ SAR snapshots along elevation 
(which gives higher elevation 
resolution but it takes much longer time to generate those $64$ snapshots). In addition, 2D depth images are the final outcomes of HwakEye, 
but in our architecture, they are only intermediate results to be used as input 
for generating 3D object shapes.  
The design of $\mathbf{G_{p2p}}$ in Stage 2 follows the idea of \cite{yuan2018pcn}, 
but unlike \cite{yuan2018pcn}, 
we use a conditional GAN architecture.
% to jointly train a generator and a discriminator. 
In addition, 
our Stage 2's generator network is designed to operate on sparse 
and only partially correct point clouds. 

\subsection{Stage 1: 3D Radar Intensity Map to 2D Depth Image}\label{sec_stage1}

%\noindent \textbf{Input data preprocessing}. 

\subsubsection{\textbf{Input and Output Representation}}

3DRIMR first pre-processes a set of 2-snapshot raw mmWave radar data
of an object by using three FFTs: \textit{range FFT, azimuth FFT, and elevation FFT}, 
to generate a 3D radar intensity map $m_r$ of the object from a particular viewpoint.  
Input $m_r$ is a $64 \times 64 \times 256$ tensor.
% after preprocessing with FFTs. 
$\mathbf{G_{r2i}}$'s output is a high-resolution 2D depth image $\hat{g}_{2d}$.
% in the stereo camera frame.
%, where the generator predicts the depth value for each pixel in the depth image.

Similar to the hybrid 3D to 2D network architecture of HawkEye \cite{HawkEye}, our $\mathbf{G_{r2i}}$  
is also a 3D-encoder-2D-decoder network. 
Different from HawkEye \cite{HawkEye}, each of our input data sample 
only contains 2 snapshots of SAR radar signals. 
%along elevation dimension, whereas the input data in HawkEye contains $40??$ snapshots in SAR operation.
Therefore our system can run much faster than HawkEye in practice.  
Recall that fewer snapshots means lower radar resolution, 
thus our network handles more challenging inputs, 
but our network can still produce high-quality outputs.
%which will be discussed in section \ref{sec_eval}.

\subsubsection{\textbf{Stage 1 training}}
%Stage 1 is a conditional GAN \cite{cGAN} based architecture. 
Given a 3D radar intensity map $m_r$,
% with only 2-snapshot RF signals, 
3DRIMR learns a generator $\mathbf{G_{r2i}}$ and a discriminator $\mathbf{D_{r2i}}$ simultaneously.
$\mathbf{D_{r2i}}$ takes $(m_r, g_{2d})$ or $(m_r, \hat{g}_{2d})$ pairs as inputs 
and learns to distinguish between $g_{2d}$ and $\hat{g}_{2d}$. 

\subsubsection{\textbf{Generator $\bf{G_{r2i}}$}}
This generator  follows a standard encoder-decoder architecture \cite{encoder-decoder},
which maps a
sparse 3D radar intensity map to a high-resolution (same level as that of the depth image taken by a stereo 
camera) 2D depth image.
As shown in Figure \ref{fig_stage_1},
its encoder consists of six 3D convolution filters, followed by Leaky-ReLU and BatchNorm layers. 
The 3D feature map is down-sampled and the number of feature channels increases after every convolution layer. 
Hence, the encoder converts a 3D radar intensity map into a low-dimensional representation. 
The decoder consists of 
%9 (in ExtraTwo) or 11 (in ExtraFour) 
nine 2D transpose convolution layers, along with ReLU and BatchNorm layers. 
The 2D feature map is up-sampled and the number of feature channels is reduced after every transpose convolution layer. Passing through the decoder network, low-dimensional vectors are finally 
converted to a high-resolution 2D depth image.

\subsubsection{\textbf{Discriminator $\bf{D_{r2i}}$}}

The discriminator takes in two inputs, 3D radar intensity maps $m_r$ and 2D depth images (either 
$g_{2d}$ or $\hat{g}_{2d}$). 
As shown in Figure \ref{fig_stage_1}, 
we design two separate networks to encode both of them into two 1D feature vectors. 
The encoder for 3D radar intensity maps is the same as $\mathbf{G_{r2i}}$'s encoder architecture. 
The encoder for 2D depth images is a typical convolutional architecture, 
which consists of 
%9 (in ExtraTwo) or 11 (in ExtraFour) 
$9$ 2D convolution layers, followed by Leaky-ReLU and BatchNorm layers.
The outputs of these two encoders are two 1D feature vectors, which are concatenated 
and then are passed into a simple convolution network with two 2D convolution layers 
and Leaky-ReLU and BatchNorm layers in between, and followed by a sigmoid function to output the final classification result. 

\subsubsection{\textbf{Skip Connection}}
We also apply skip connection in $\mathbf{G_{r2i}}$ 
to help avoid the gradients vanishing problems in deep architectures, 
and to make the best use of the input data since the range or depth information of an input 3D intensity map 
can be directly extracted and passed to higher layers. 
We choose $8$ max values of a 3D intensity map along range dimension to form a $8$-channel 2D feature map. 
%that is because only choose one most largest value is not stable. 
Then, we concatenate this feature map with the feature map produced by the $6$-th layer of the decoder, 
and they together are passed through the remaining decoder layers.

\subsubsection{\textbf{Loss Functions}}
Our architecture trains $\mathbf{D_{r2i}}$ to minimize discriminator loss $\mathcal{L}_{\mathbf{D_{r2i}}}$, 
and train $\mathbf{G_{r2i}}$ to minimize generator loss $\mathcal{L}_{\mathbf{G_{r2i}}}$ simultaneously. 
%The loss function of discriminator 

$\mathcal{L}_{\mathbf{D_{r2i}}}$ is 
the mean MSE (Mean Square Error) of $\mathbf{D_{r2i}}$'s prediction error
when the input of includes $(m_r, \hat{g}_{2d})$ and $(m_r, g_{2d})$.

$\mathcal{L}_{\mathbf{G_{r2i}}}$ is a weighted sum of vanilla GAN loss $\mathcal{L}_{GAN}$, 
$\mathcal{L}_1$ loss between $\mathbf{G_{r2i}}$'s prediction and ground truth, 
and a perceptual loss $\mathcal{L}_{p}$. 
The perceptual loss is calculated by a pre-trained neural network VGG\cite{VGG} 
on $\mathbf{G_{r2i}}$'s prediction and ground truth.
\begin{equation}\label{LG}
\mathcal{L}_{\mathbf{G_{r2i}}} = \mathcal{L}_{GAN} (\mathbf{G_{r2i}}) + \lambda_1 \mathcal{L}_{1} (\mathbf{G_{r2i}}) + \lambda_p \mathcal{L}_{p} (\mathbf{G_{r2i}})
\end{equation}
%3DRIMR Stage 1's training objective is to reduce $\mathcal{L}_{\mathbf{G_1}}$, 
%a weighted combination of three terms as shown in Eqn. (\ref{LG}). 
Note that $\lambda_1$ and $\lambda_p$ are two hand-tuned relative weights, and in our simulations, we find that it performs well when the values of them are 1000 and 20 respectively.

\subsection{Stage II:  Multi-View 2D Depth Images To 3D Point Clouds}\label{sec_stage2}

As shown in Figure \ref{fig_stage_2}, 
this stage consists of another conditional GAN based neural network $\mathbf{G_{p2p}}$
that generates a point cloud of an object with continuous and smooth contour $\hat{P}_r$, 
based on $P_r$, a union of $k$ separate coarse point clouds 
observed from $k$ viewpoints of the object $\{P_{r,i} | i=1,...,k\}$, 
which are the outputs from Stage 1. We choose $k=4$
in our experiments.  
Note that $P_r$ may contain very noisy or even incorrect points due to the prediction errors
in Stage 1. 

Both input $P_r$ and output $\hat{P}_r = \mathbf{G_{p2p}}(P_r)$ 
are 3D point clouds represented as $n \times 3$ matrices, 
and each row is the 3D Cartesian coordinate $(x, y, z)$ of a point. 
Note that the input and output point clouds do not necessarily have the same number of points.
Generator $\mathbf{G_{p2p}}$ is an encoder-decoder network in which
an encoder first transforms an input point cloud $P_r$ into a k-dimensional feature vector, 
and then a decoder outputs $\hat{P}_r$.
The discriminator  $\mathbf{D_{p2p}}$ of this stage takes $(P_r, P_{true})$ 
or $(P_r, \hat{P}_r)$ 
pairs and produces a score to distinguish between them. 

\subsubsection{\textbf{Generator $\mathbf{G_{p2p}}$}}

%We leverage the partial architecture of PCN \cite{yuan2019pcn}, which is also an encoder-decoder based network.
The generator also adopts an encoder-decoder structure. 
The encoder has two stacked PointNet \cite{qi2016pointnet} blocks. 
This point feature encoder design follows similar designs in 
PointNet \cite{qi2016pointnet} and PCN \cite{yuan2018pcn}, 
characterized by permutation invariance and tolerance to noises. 

The \textbf{first block} of $\mathbf{G_{p2p}}$ takes the input $P_r$, 
and uses a shared multi-layer perceptron (MLP) to produce a high-dimensional local feature $f_i (i=1, ..., n)$  for each point $p_i$ in $P_r$, 
and hence we can obtain a local feature matrix $F_{P_r}$ with each row being
the local feature of the corresponding point.
Then, it applies a point-wise maxpooling on $F_{P_r}$ and extracts a high-dimensional global feature vector $g_{P_r}$.
To produce a complete point cloud for an object, we need both local and global features, 
hence, the encoder concatenates the global feature $g_{P_r}$ with each of the point features 
$f_i$ and form another matrix $F'_{P_r}$.
The \textbf{second block} also consists of a shared MLP and point-wise maxpooling layer, 
which takes $F'_{P_r}$ as input and produces the final global feature $g_f$.

The decoder consists of 3 fully connected layers with proper non-linear activation layers
 and normalization layers. 
It expands $g_f$ to a $3m$ dimensional vector, and then reshapes it to a $m \times 3$ matrix. 
The matrix has $m$ rows and each row represents the Cartesian coordinate $(x, y, z)$ of a point of the predicted point cloud.
The fully-connected decoder design is good at predicting a sparse set of points which represents the global 
geometry of a shape. 

\subsubsection{\textbf{Discriminator $\mathbf{D_{p2p}}$}}

Note that the number of points of input $P_r$ and output $\hat{P}_r$ and ground truth $P_{true}$ 
can be different, and the points of a point cloud are unordered. 
Therefore we need to design a discriminator with two-stream inputs.
As shown in Figure \ref{fig_stage_2}, 
these two inputs pass through the same architecture as used in $\mathbf{G_{p2p}}$ 
and are converted into their own final global feature, which are concatenated and fed into 2 fully connected 
layers to produce a score, which is used to indicate whether the input is real or generated point cloud.

\subsubsection{\textbf{Loss Function}}

For this GAN based network, we train $\mathbf{D_{p2p}}$ to minimize $\mathcal{L}_{\mathbf{D_{p2p}}}$, 
and train $\mathbf{G_{p2p}}$ to minimize $\mathcal{L}_{\mathbf{G_{p2p}}}$ simultaneously. 
$\mathcal{L}_{\mathbf{D_{p2p}}}$ is calculated in the same way as described in Stage 1, 
but $\mathcal{L}_{\mathbf{G_{p2p}}}$ is different due to the characteristics of point clouds.
$\mathcal{L}_{\mathbf{G_{p2p}}}$ is a weighted sum, 
consisting of $\mathcal{L}_{GAN} (\mathbf{G_{p2p}})$, 
Chamfer loss $\mathcal{L}_{cf}$  
between predicted point clouds and the ground truth, 
and IoU loss $\mathcal{L}_{iou} (\mathbf{G_{p2p}})$. 
Chamfer distance \cite{yuan2018pcn} is defined as:
\begin{equation}
\small{
	d_{cf}(S_1, S_2) = 
\frac{1}{|S_1|} \sum_{x \in S_1} \mathop{min}\limits_{y \in S_2} \lVert x-y \rVert_{2} 
+ \frac{1}{|S_2|} \sum_{y \in S_2} \mathop{min}\limits_{x \in S_1} \lVert y-x \rVert_{2}
}
\end{equation}
%Let's use $P_{pred}$ and $P_{gt}$ to represent predicted point cloud the ground truth respectivelly. 
In our case, Chamfer loss is calculated as:
\begin{equation}
\mathcal{L}_{cf} (\mathbf{G_{p2p}}) = d_{cf}(\hat{P}_r, P_{true})
\end{equation}
We define IoU of two point clouds as the ratio of the number of shared common points 
over the number of points of the union set of both two point clouds. 
First, we voxelize the space,
% with each voxel’s edge length being $r$, 
and all the points in the same voxel will be treated as the same point.
We define the number of voxels occupied by $\hat{P}_r$ and $P_{true}$ as $\hat{V}_r$ and $V_{true}$ respectively.
Then, the $IoU$ can be calculated as:
\begin{equation}
IoU = \frac{\hat{V}_r \cap  V_{true}}{\hat{V}_r \cup V_{true} + \epsilon }
\end{equation}
Note that $\epsilon$ is a very small number such as $10^{-6}$ to avoid dividing by zero.
Since IoU is within range 0 and 1, we define our IoU loss as:
\begin{equation}
\mathcal{L}_{IoU} (G) = 1 - IoU
\end{equation}
$\mathcal{L}_{\mathbf{G_{p2p}}}$ is given by Eqn. (\ref{eqn_L_G_p2p}), 
and $\lambda_{d_{cf}}$ and $\lambda_{iou}$ are also need to be hand-tuned. 
In our experiments, we set the values of them to 100 and 10 respectively.
\begin{equation}\label{eqn_L_G_p2p}
\mathcal{L}_{\mathbf{G_{p2p}}} = \mathcal{L}_{GAN} (\mathbf{G_{p2p}}) + \lambda_{d_{cf}} \mathcal{L}_{cf} (\mathbf{G_{p2p}}) 
+ \lambda_{iou} \mathcal{L}_{iou} (\mathbf{G_{p2p}})
\end{equation}

\section{Implementation and Experiments}\label{sec_imp}

We have implemented 3DRIMR architecture and conducted experiments to test it with real and synthesized mmWave  data. 

\subsection{Datasets}

To the best of our knowledge, there is no publicly available dataset of mmWave radar sensing, 
therefore we have conducted experiments to collect real data,  
%via TI's IWR6843ISK radar sensor \cite{iwr6843}. 
and since real data collection is time consuming, 
we have also augmented our dataset via synthesizing mmWave data as done in \cite{HawkEye}.

\noindent \textbf{Real radar data collection.} 
We collected real radar data with Synthetic Aperture Radar (SAR) operation 
using an IWR6843ISK \cite{iwr6843} sensor and a data capture card DCA1000EVM \cite{dca1000evm}. 
%configured by the mmWave studio \cite{??}. 
To generate a full-scale 3D energy intensity data as a baseline, 
we first conducted SAR radar scans with a $24\times 64$ virtual antenna array of the sensor,
by sliding the sensor horizontally and vertically through a customized slider. 
%$3$ positions horizontally\footnote{The sensor has $8$ virtual antennas horizontally at each position.}
%and $64$ positions vertically. 
Note that the sensor's each scan has a range or depth of $256$ units.  
We expanded a $24\times 64\times 256$ 3D data cube (obtained via a SAR scan) to a full-scale 
$64\times 64 \times 256$ 3D data cube, 
which can be regarded as consisting of $64$ snapshots stacked vertically, 
with each snapshot being a data plane of size $64\times 256$.

Note that the full-scale data is only used as a base reference in 
validating our 3DRIMR deep learning system's effectiveness. 
When training and testing 3DRIMR, 
each actual input data sample has only $2$ snapshots out of $64$ snapshots. 
Getting $2$ snapshot data is a much faster process than 
getting full-scale $64$ snapshot data.

\noindent \textbf{Synthesized radar data.} 
Similar to \cite{HawkEye}, we used 3D CAD models of cars \cite{fidler20123d}
to generate synthesized radar signals.
% and 2D depth images. 
%we convert CAD models in point-cloud format using FreeCAD \cite{??}, then synthesize the radar signals and 2D 
%depth images in MATLAB. 
We first generated 3D point clouds based on those CAD models, 
and translated and rotated them to simulate different scenarios. 
Then we selected points in each point cloud as radar signal reflectors. 
Next we simulated received radar signals in a receiver antenna array based on our radar configuration.

\noindent \textbf{Ground truth 2D depth images and point clouds.}
For real data, ground truth depth images were obtained via a ZED mini camera \cite{zed}. 
%that captures 2D depth images from the scenes. 
Note that ZED mini camera is widely used in mobile robots and VR sets.  
For synthesized data, based on 
%the selected camera reflectors of 
the derived point clouds of CAD models, we generated ground truth depth images after perspective projection 
with appropriate camera settings and viewpoints.
Similarly we generated ground truth 3D point clouds. 
%using the same method as we use the ground truth depth images. 

\noindent \textbf{Generating 3D radar energy intensity data.} 
For both real and synthesized radar data, we performed FFT along all three dimensions (i.e., azimuth $\phi$ and elevation $\theta$ and range $r$).
Note that a radar 3D data cube is measured in degree along azimuth $\phi$ and 
elevation $\theta$ dimension. We converted a data cube into Cartesian coordinate system so that 
it matched the coordinate system of the depth camera. 
This is different from \cite{HawkEye} which directly used original data in 
spherical coordinate system (hence introduced additional errors).

\subsection{Model Training and Testing}

We conducted experiments of 3DRIMR on cars (large objects with 
average size of $445cm \times 175cm \times 158cm$), and a L-shaped box (a small object with size of $95cm \times 73cm\times 59cm$).

\subsubsection{\textbf{Cars}} 
%Inspired by \cite{HawkEye}, we generated synthesized data of cars to test 3DRIMR. 
In Stage 1 of 3DRIMR, 
the training dataset consisted of $8$ different categories of car models,
and for each car model, we collected data of $300$ different car orientations from $4$ views, 
so in total $300 \times 4 \times 8 = 9600$ data samples were used in training. 
We tested the model using another $6400$ data samples. 
%Note that each data sample consisted of
%a ground truth 2D depth image and a 3D radar intensity map.
In Stage 2 of 3DRIMR, we used the 2D depth images generated in Stage 1 to form a dataset of coarse point clouds,
which included $1600$ point clouds with $200$ point clouds of each car model. 
We trained Stage 2's networks for $200$ epochs using $1520$ point clouds with batch size $4$.
%The learning rate for the first 100 epochs is $2 \times 10^{-4}$ and linearly decreases to 0 in the rest 100 epochs.
Then we tested the generator network using the remaining $80$ point clouds.

\subsubsection{\textbf{L-Box}} 
%\noindent \textbf{Lbox Reconstruction} 
We collected radar data for a L-shaped box placed in our lab and let a mobile robot carry a tall flag 
and move around in the room to simulate humans walking around, which dynamically re-directed radar signals. 
We also augmented the dataset by generating synthesized data for the same L-Box via CAD models placed in the same settings as used in the real data collection.
In Stage 1 of 3DRIMR, we trained the networks based on $320$ real data samples and $4800$ synthesized data samples 
with batch size $4$ for $200$ epochs. 
We tested the model using $80$ real data samples and $6400$ synthesized data samples.
To train the networks of 3DRIMR's Stage 2, 
we used $10$ point clouds generated by Stage 1's output of real data and $1500$ point clouds generated by Stage 1's output of synthesized data.
We used the rest $110$ data samples ($10$ from real and $100$ from synthesized data) to test the trained networks.

\begin{figure}[htb!]
	\centerline{
		\begin{minipage}{1.8in}
			\begin{center}
				\setlength{\epsfxsize}{1.8in}
				\epsffile{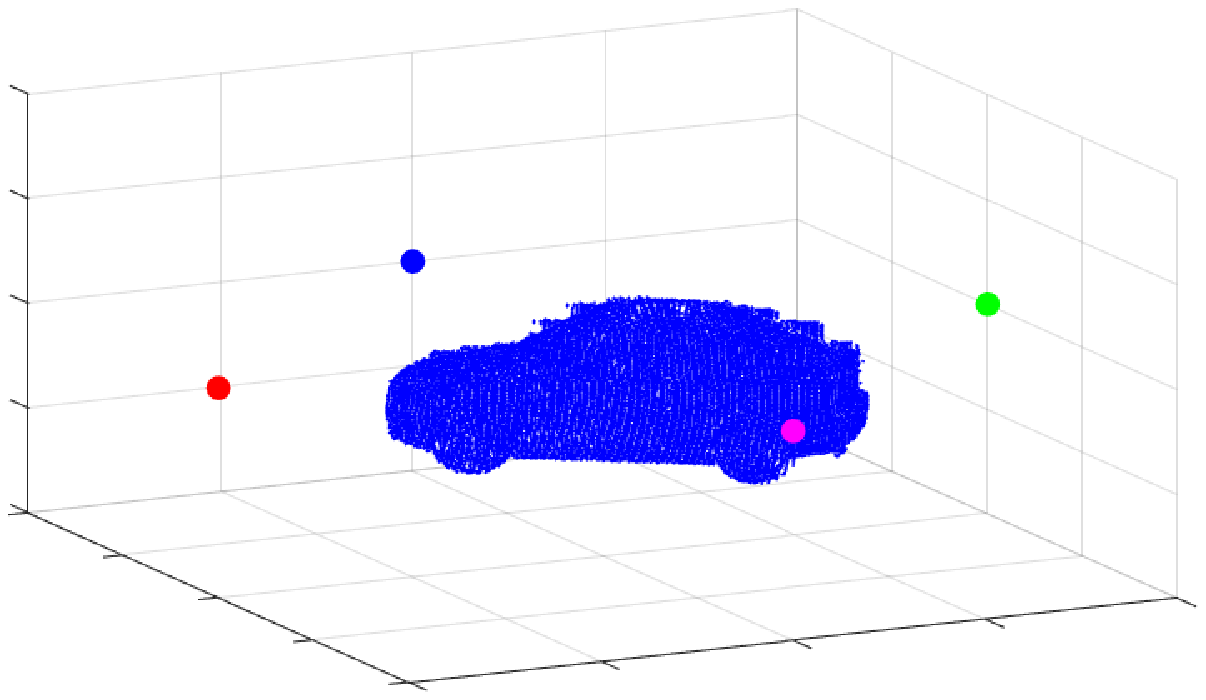}\\
				{}
			\end{center}
			%\caption{An example scene of car and radar/camera poisitions (colored dots in the figure).}
			\caption{An example scene of car.}
			\label{fig_car_ex}
		\end{minipage}	
		\begin{minipage}{1.8in}
			\begin{center}
				\setlength{\epsfxsize}{1.8in}
				\epsffile{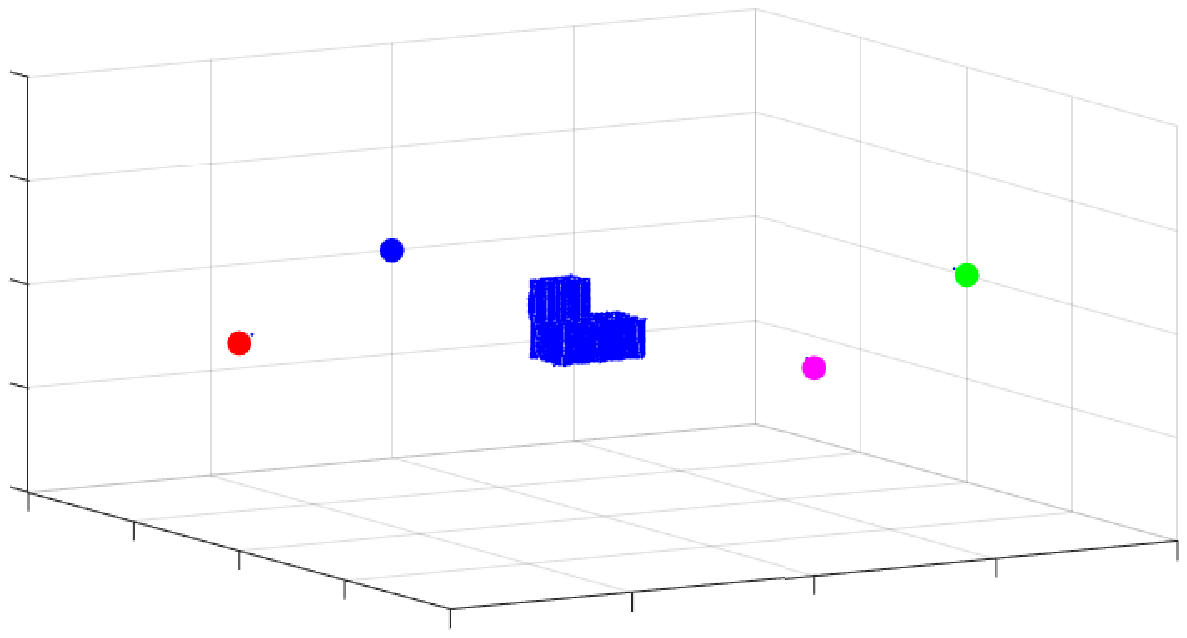}\\					
				{}
			\end{center}
			%\caption{An example scene of L-box and cameras (the colored dot in the figure).}
			\caption{An example scene of L-box.}
			\label{fig_Lbox_ex}
		\end{minipage}
	}
\end{figure}

\begin{figure*}[htb!]
	%\centerline{
	% cam1
	\begin{minipage}{7.0in}
		\begin{center}
			\setlength{\epsfxsize}{1.6in}
			\epsffile{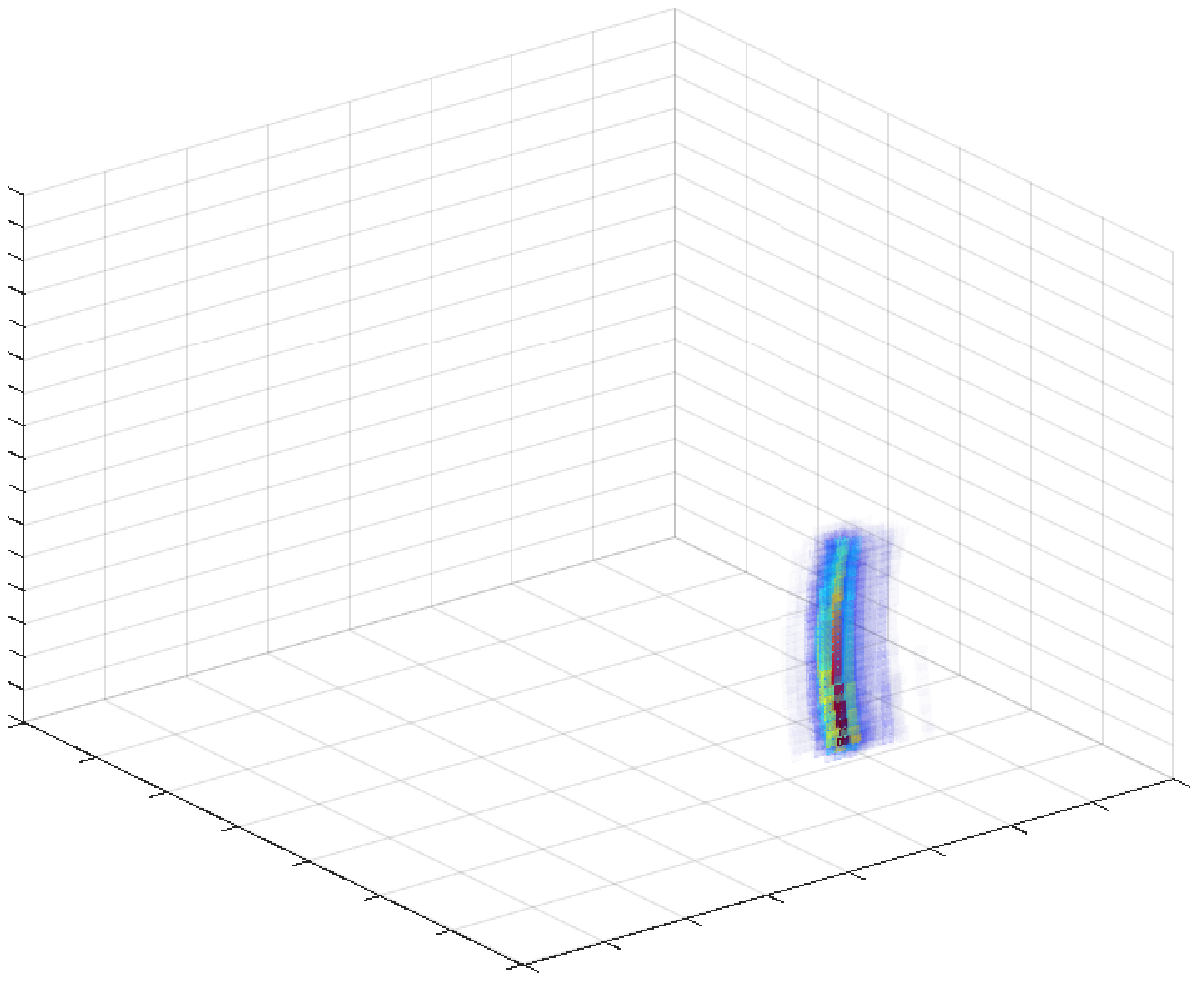}
			\setlength{\epsfxsize}{1.6in}
			\epsffile{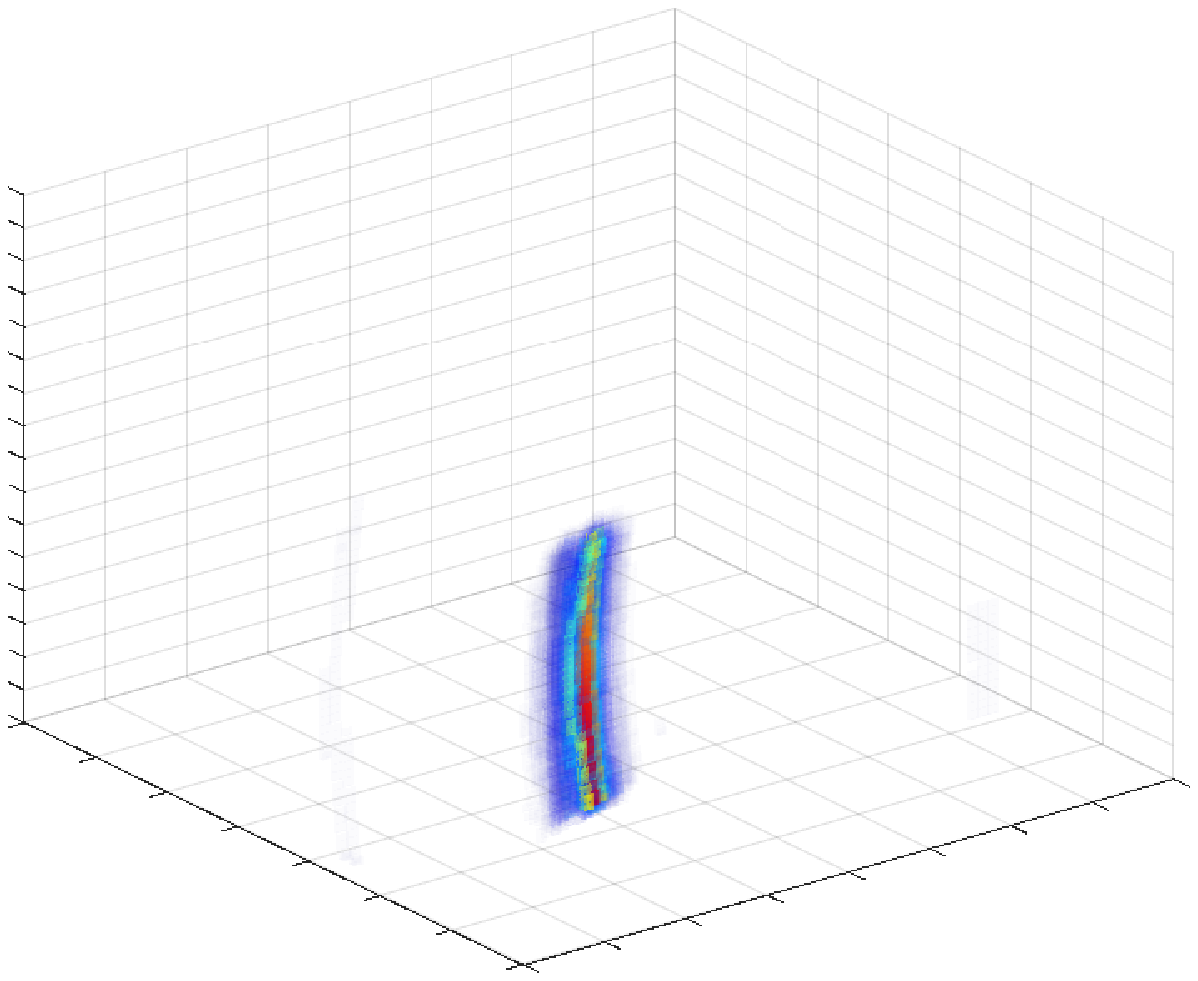}	
			\setlength{\epsfxsize}{1.6in}
			\epsffile{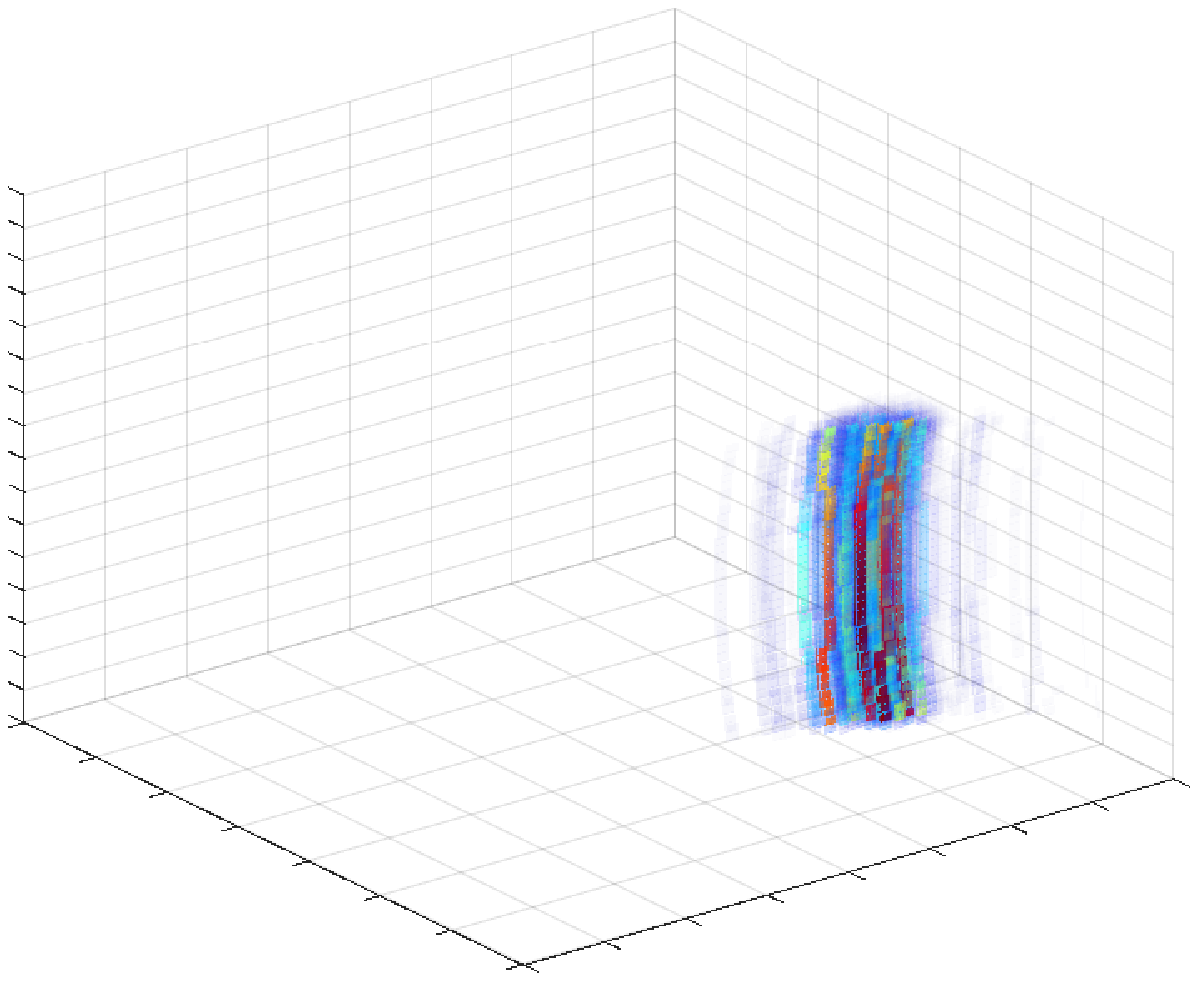}
			\setlength{\epsfxsize}{1.6in}
			\epsffile{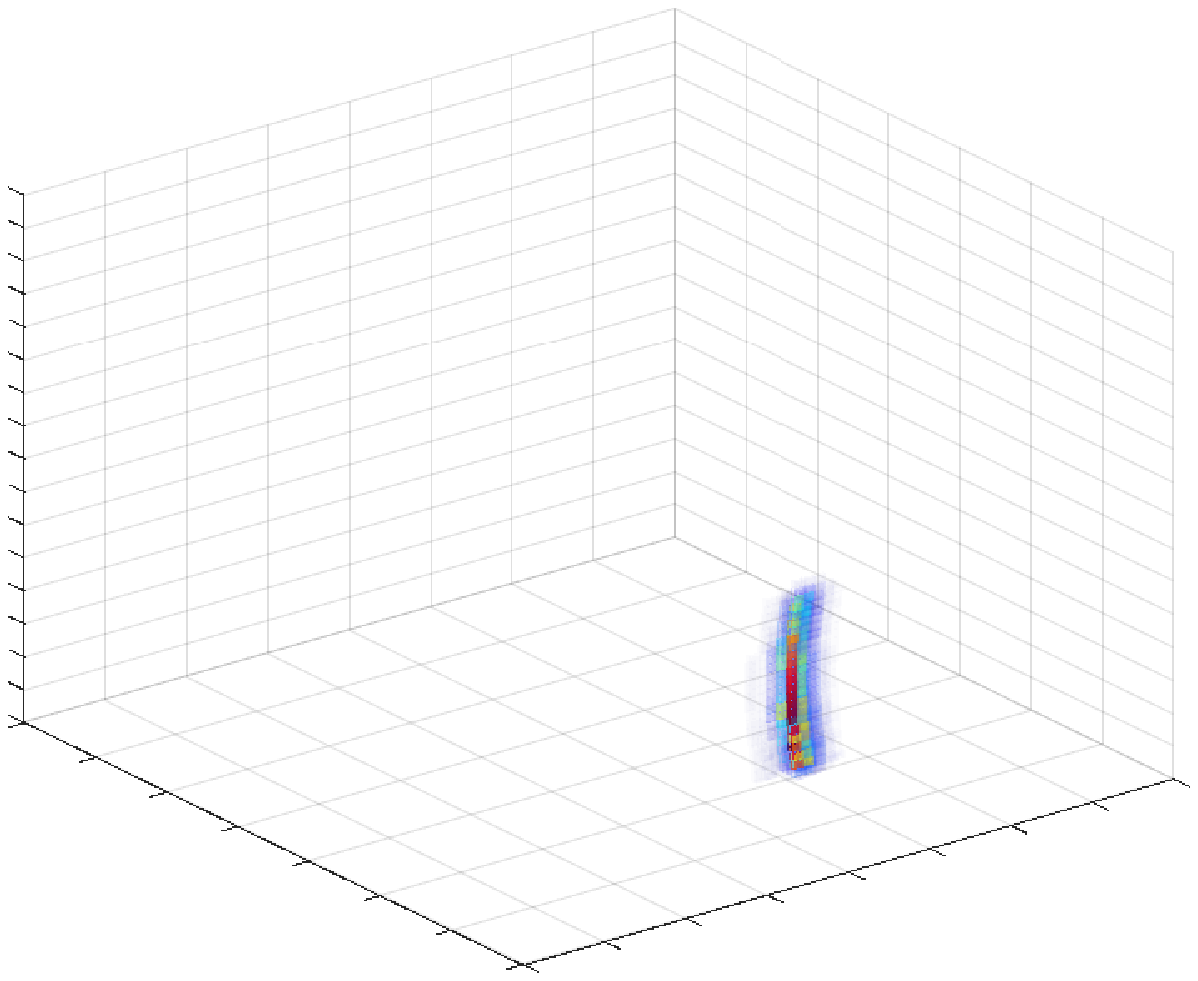}
		\end{center}
	\end{minipage}\label{fig_Lbox91_cam4_radar}\\
	
	\begin{minipage}{7.0in}
		\begin{center}
			\setlength{\epsfxsize}{1.6in}
			\epsffile{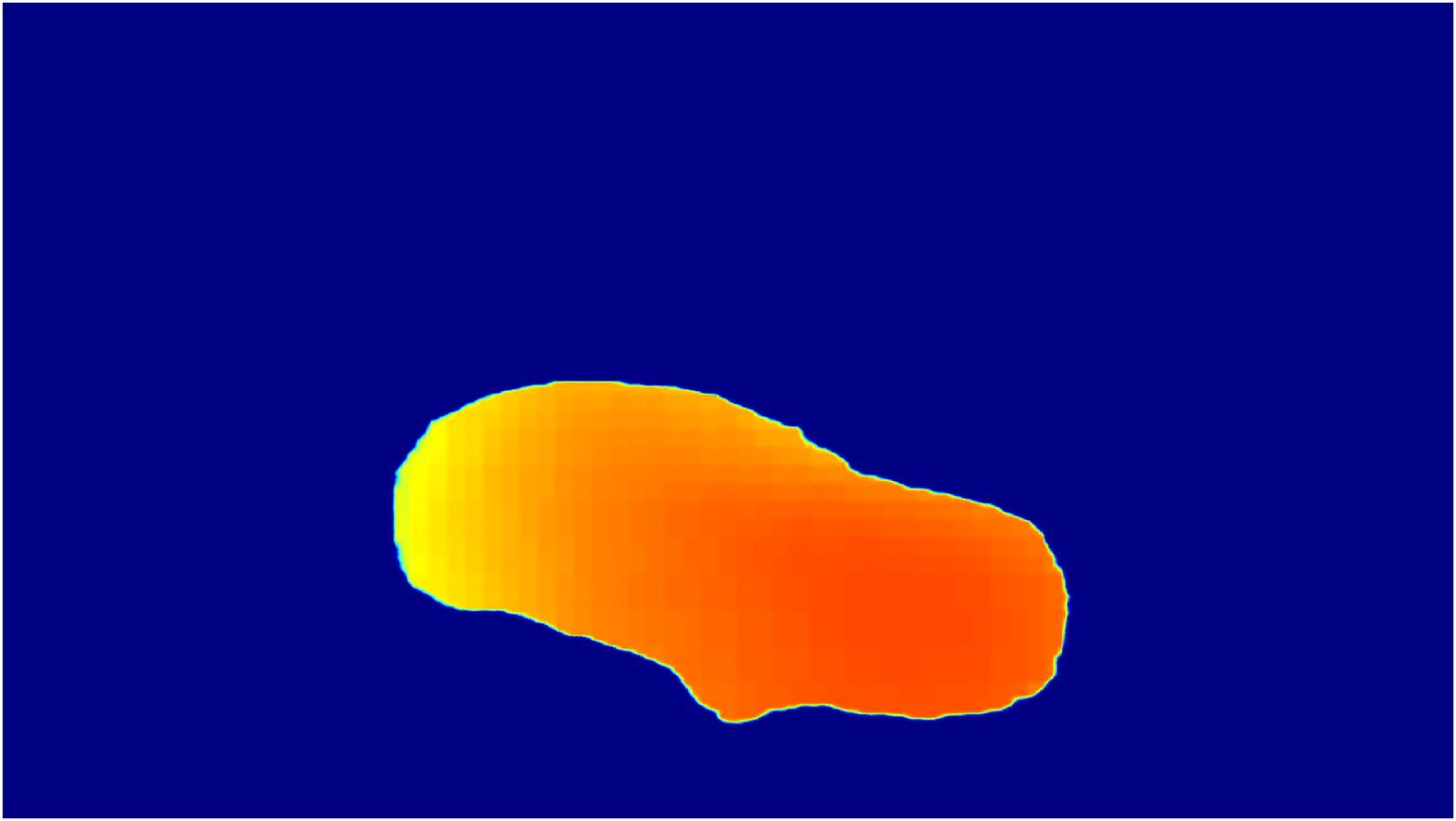}
			\setlength{\epsfxsize}{1.6in}
			\epsffile{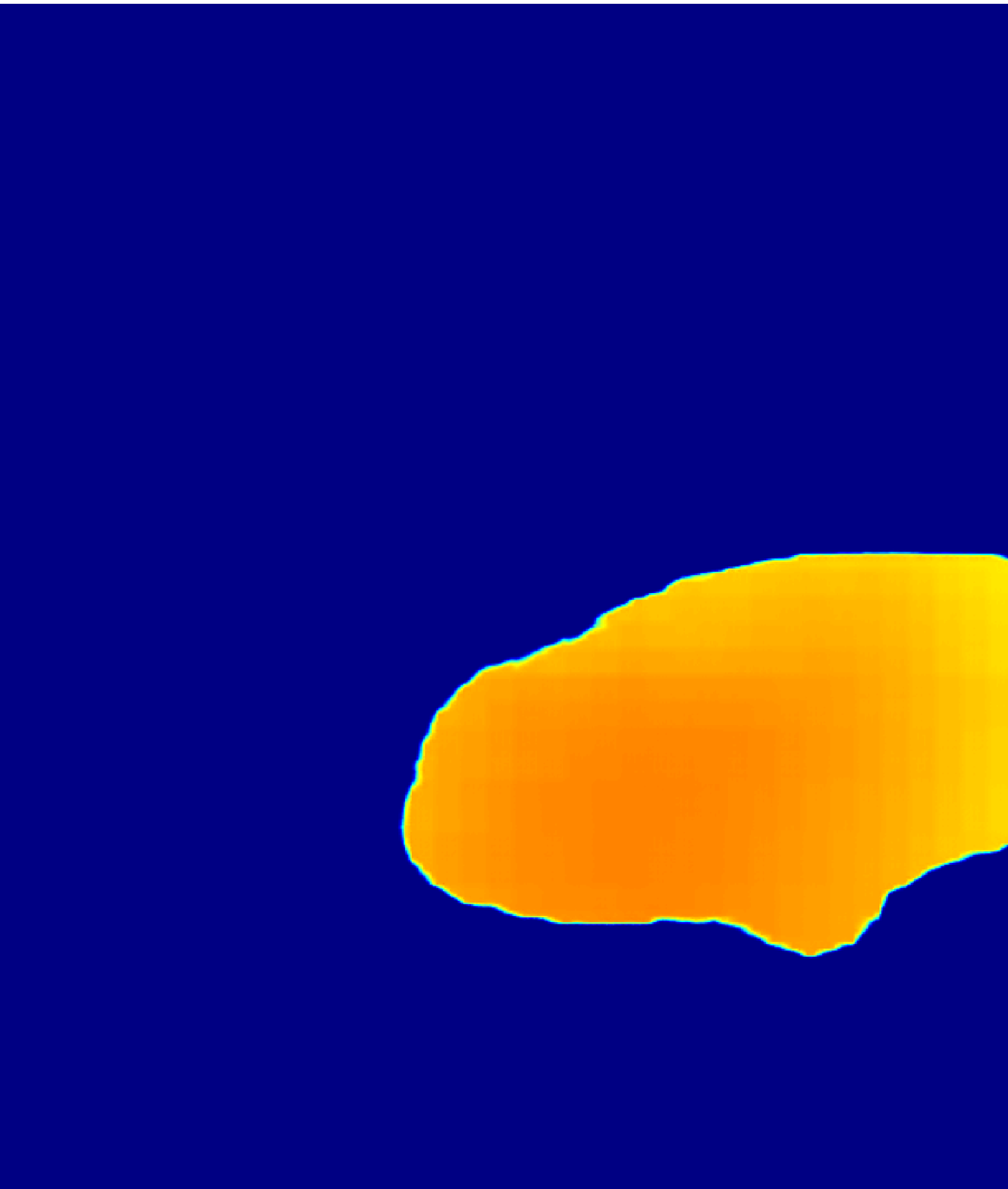}			
			\setlength{\epsfxsize}{1.6in}
			\epsffile{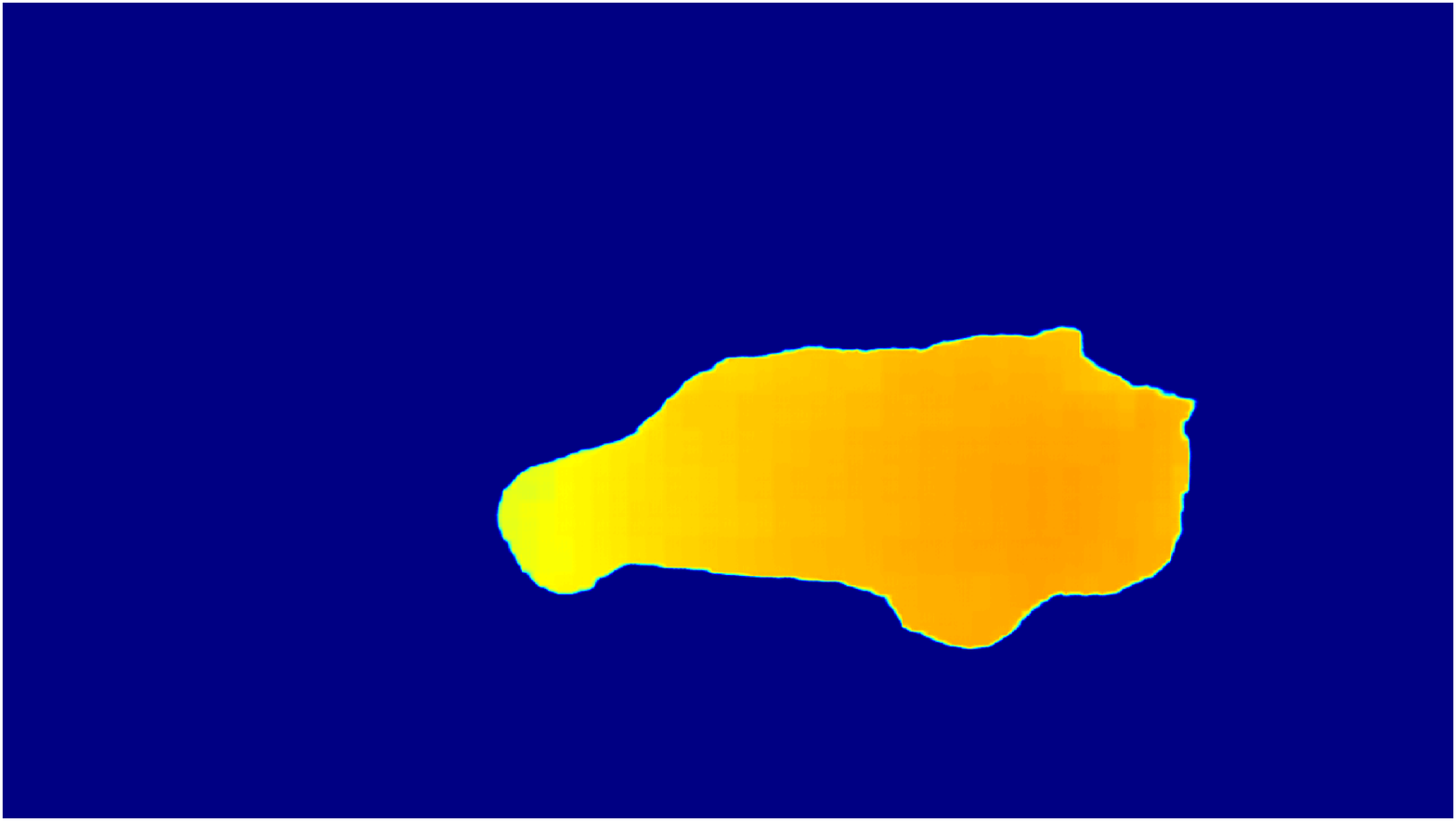}
			\setlength{\epsfxsize}{1.6in}
			\epsffile{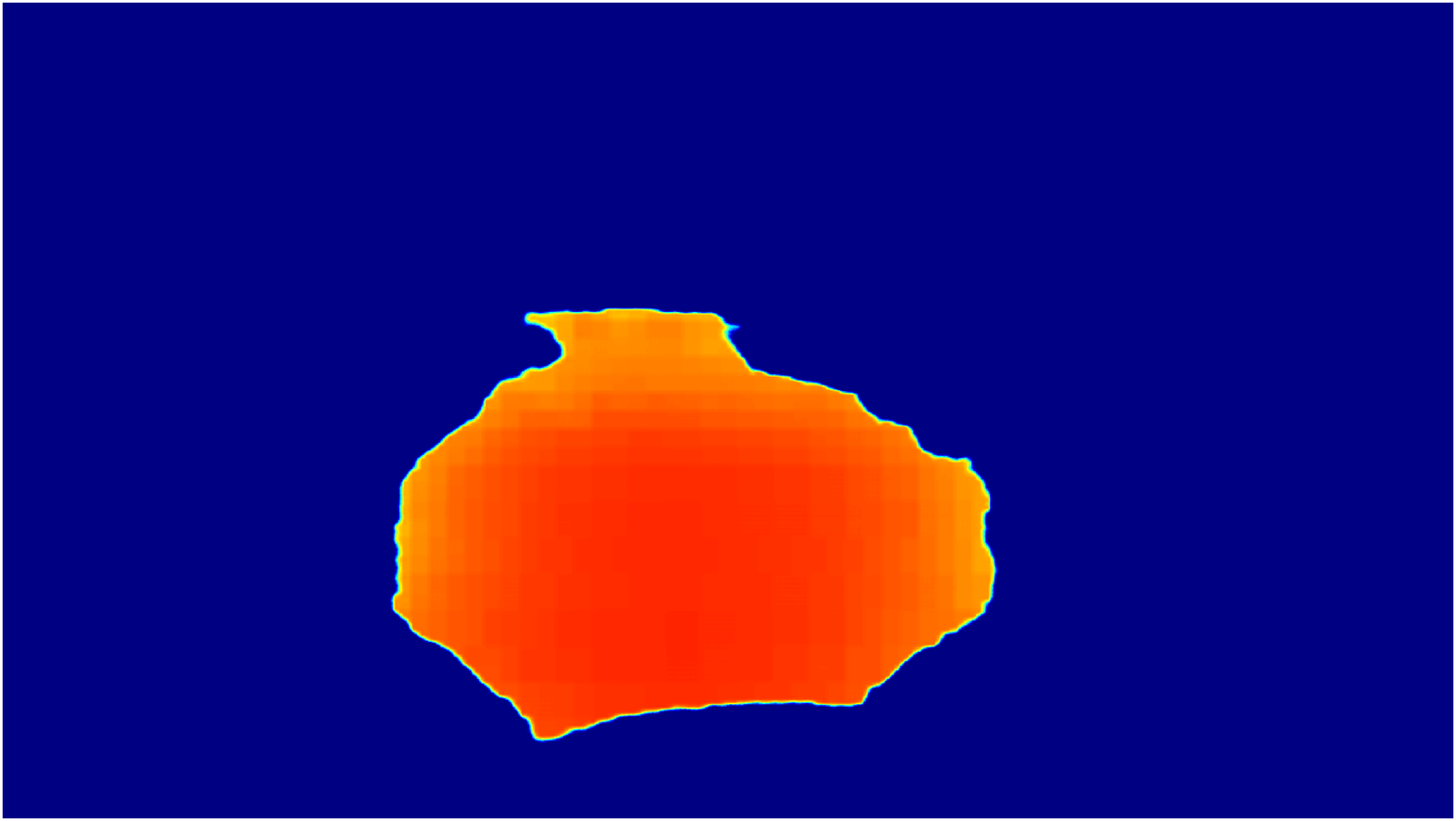}				
		\end{center}
	\end{minipage}\vspace{0.05in}\label{fig_Lbox91_cam4_fake-depth}\\
	\begin{minipage}{7.0in}
		\begin{center}
			\setlength{\epsfxsize}{1.6in}
			\epsffile{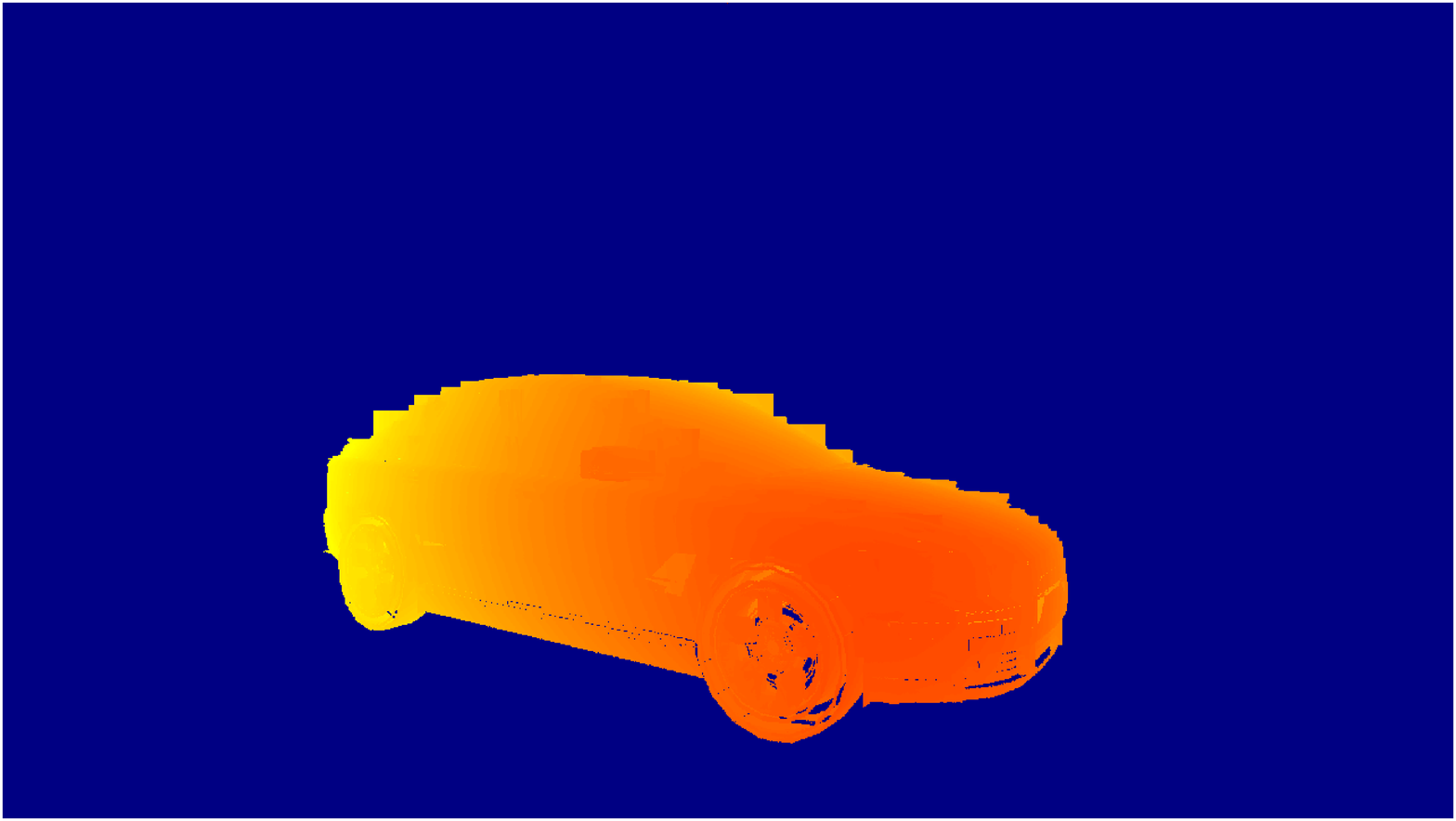}
			\setlength{\epsfxsize}{1.6in}
			\epsffile{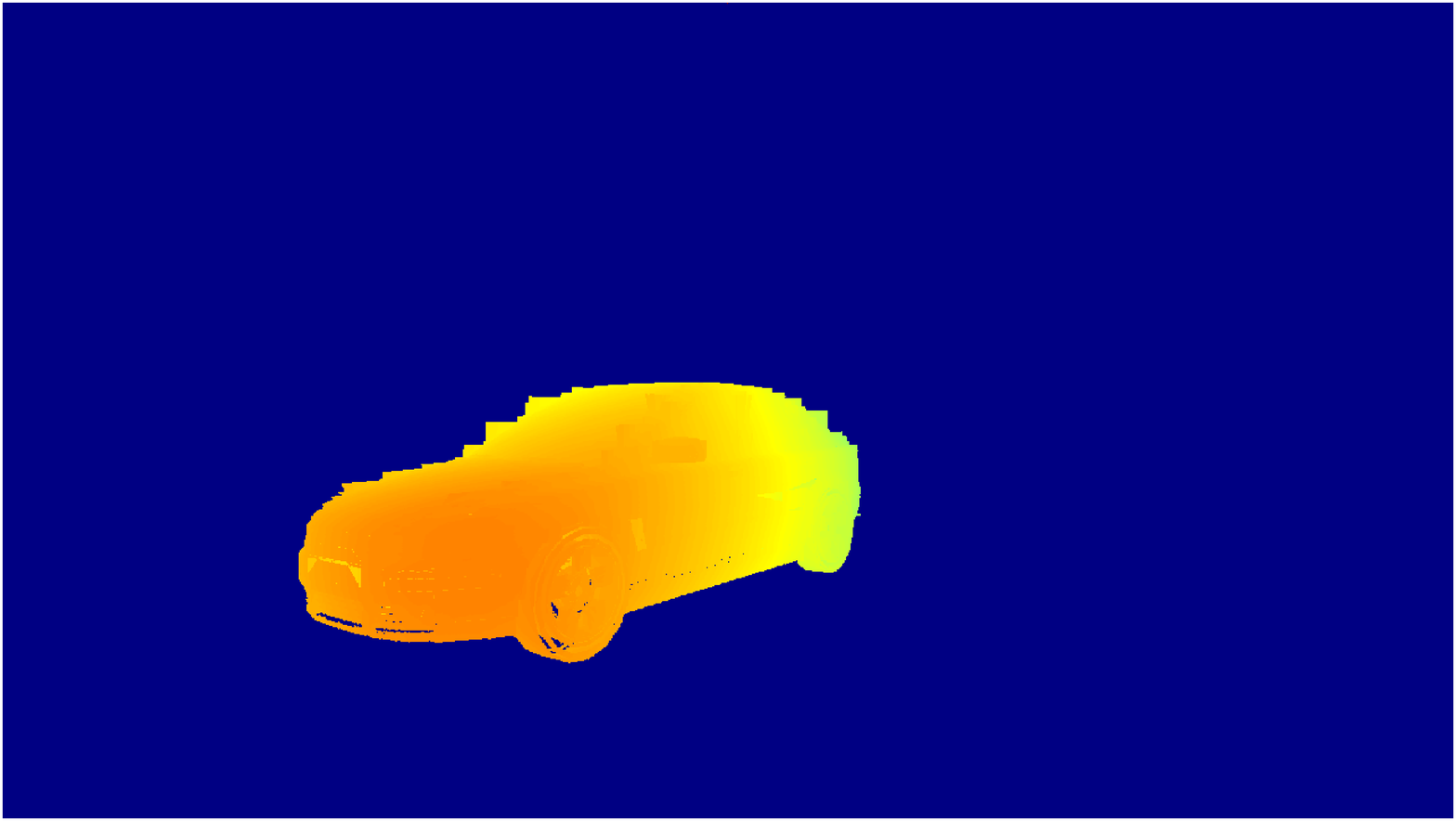}
			\setlength{\epsfxsize}{1.6in}
			\epsffile{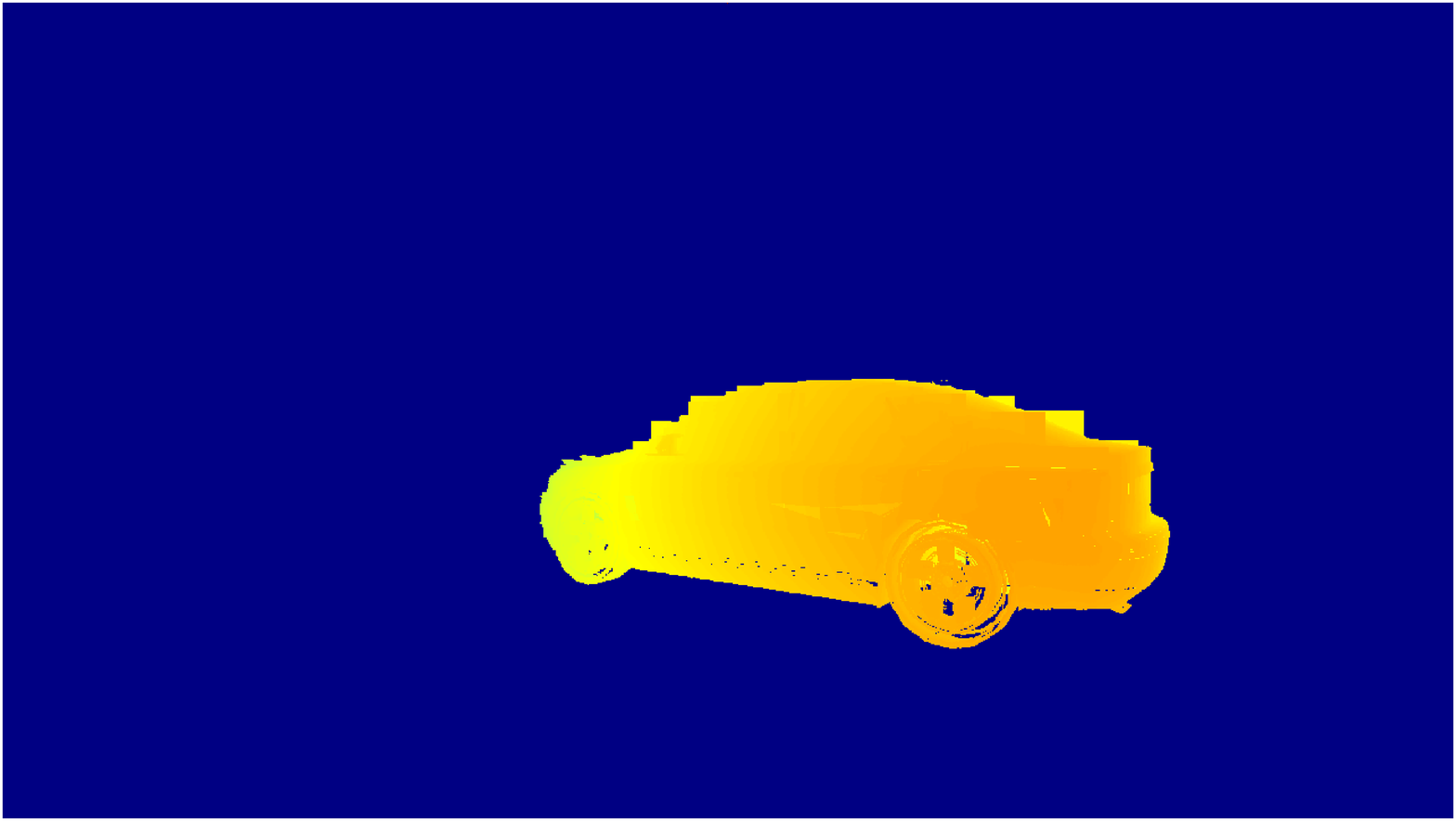}			
			\setlength{\epsfxsize}{1.6in}
			\epsffile{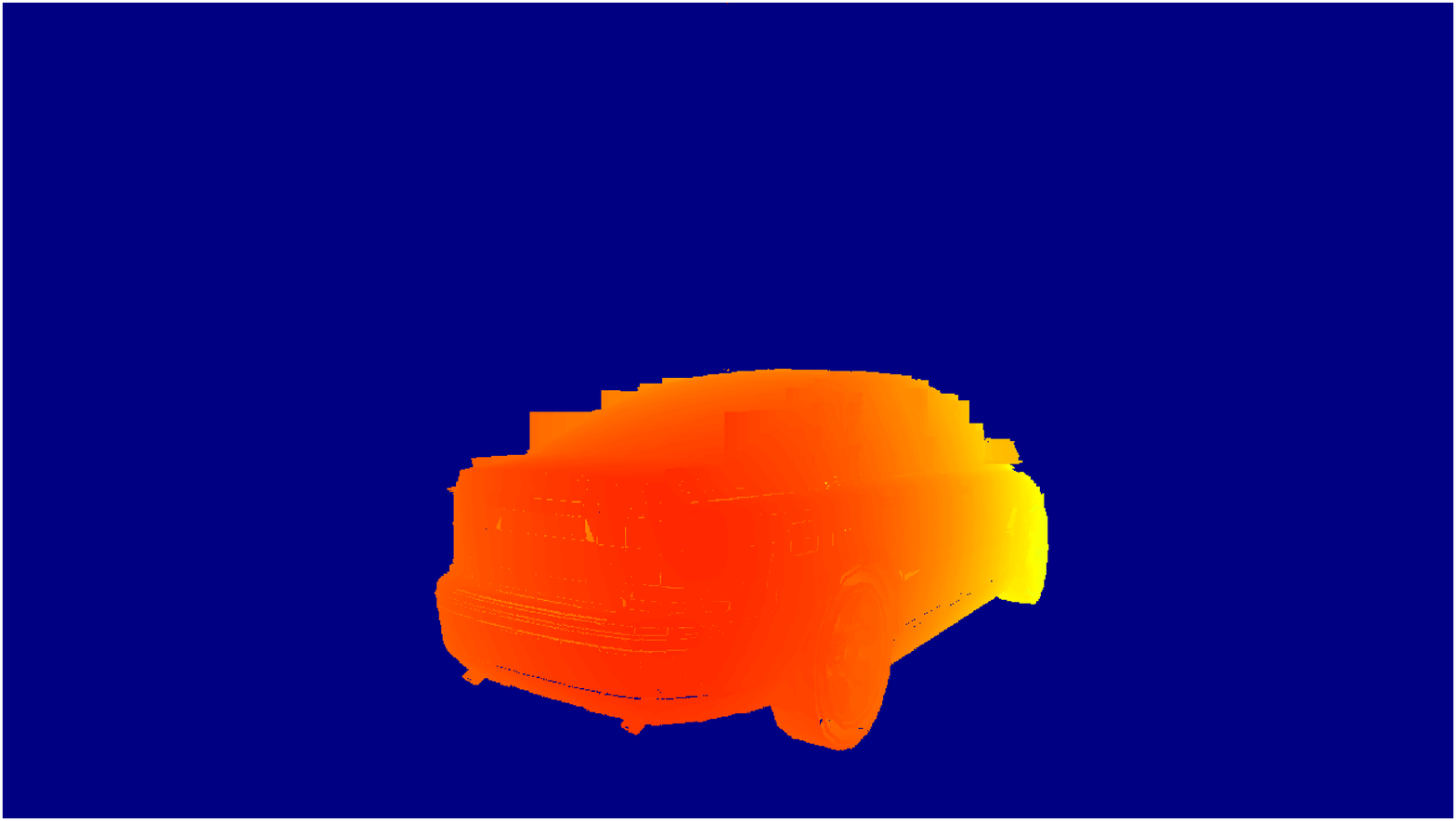}					
		\end{center}
		\label{fig_Lbox91_cam4_real-depth}
	\end{minipage}
	%}
	\caption{3DRIMR Stage 1's qualitative performance in car experiments. The 1st row shows the 3D radar intensity data from 2 snapshots only. The 2nd row shows the outputs from 3DRIMR's Stage 1. The 3rd row shows the ground truth depth images.}
	\label{fig_stage1_car}
\end{figure*}

\begin{figure*}[htb!]
	%\centerline{
	% cam1
	\begin{minipage}{7.2in}
		\begin{center}
			\setlength{\epsfxsize}{1.7in}
			\epsffile{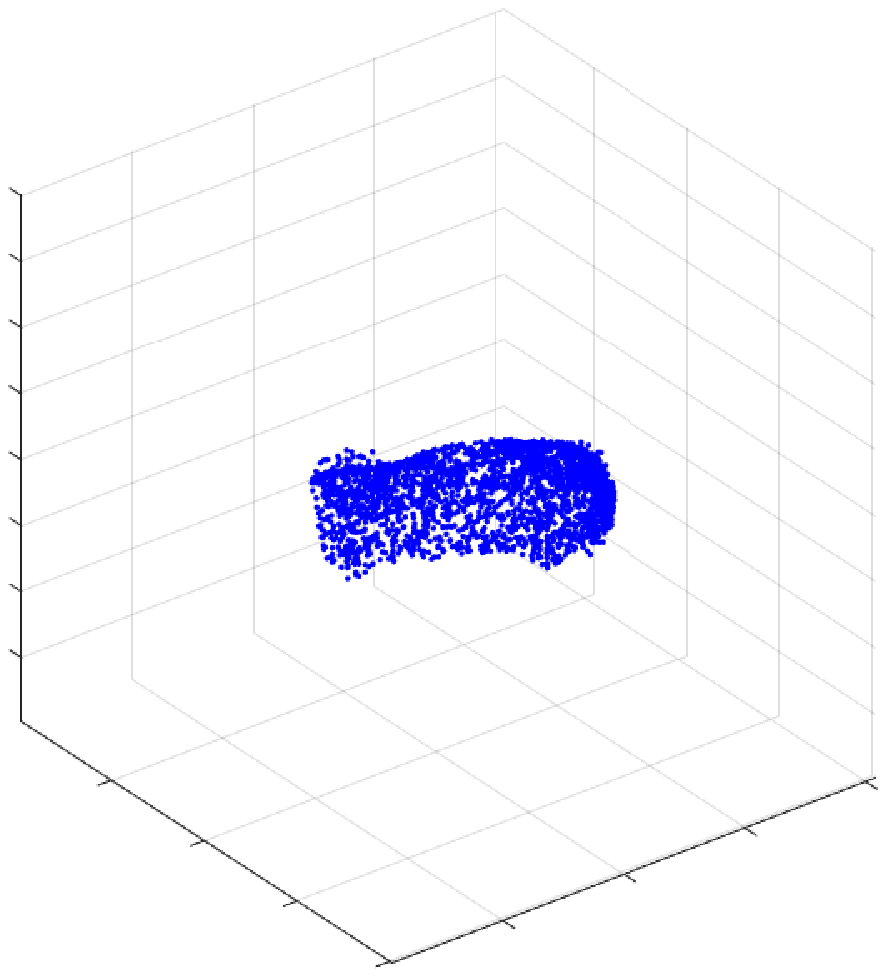}
			\setlength{\epsfxsize}{1.7in}
			\epsffile{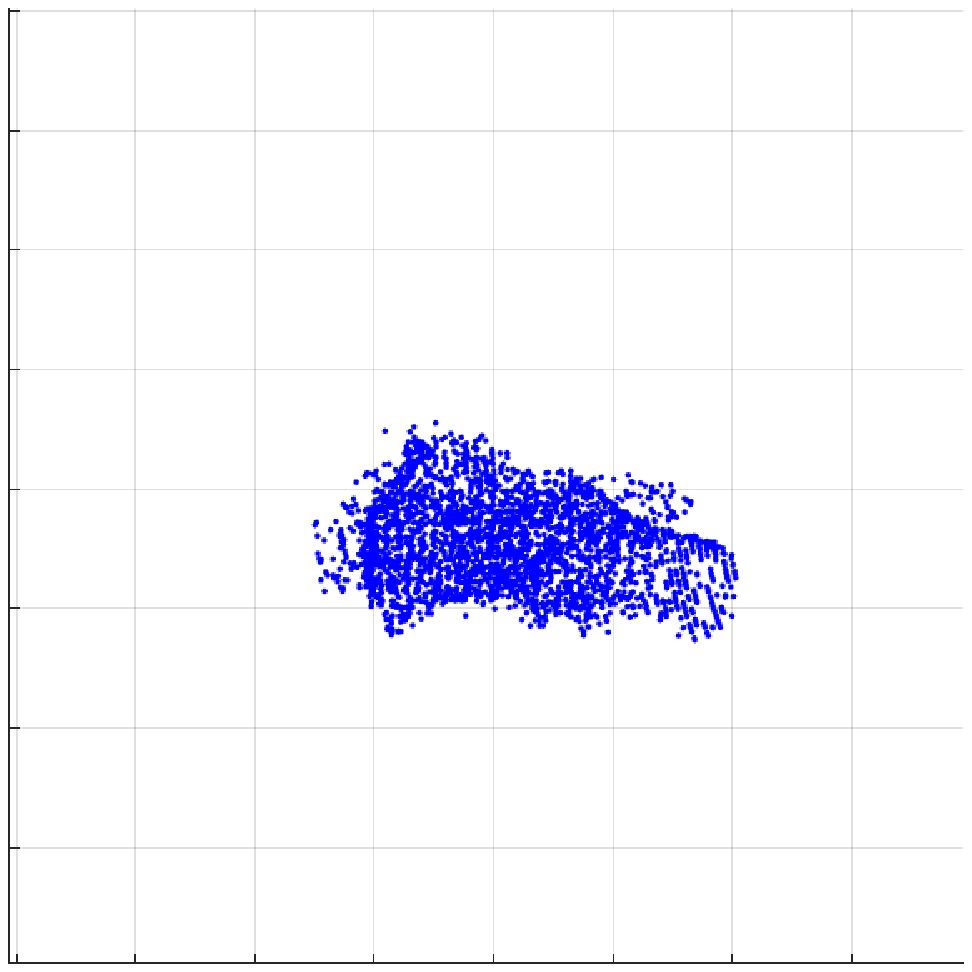}	
			\setlength{\epsfxsize}{1.7in}
			\epsffile{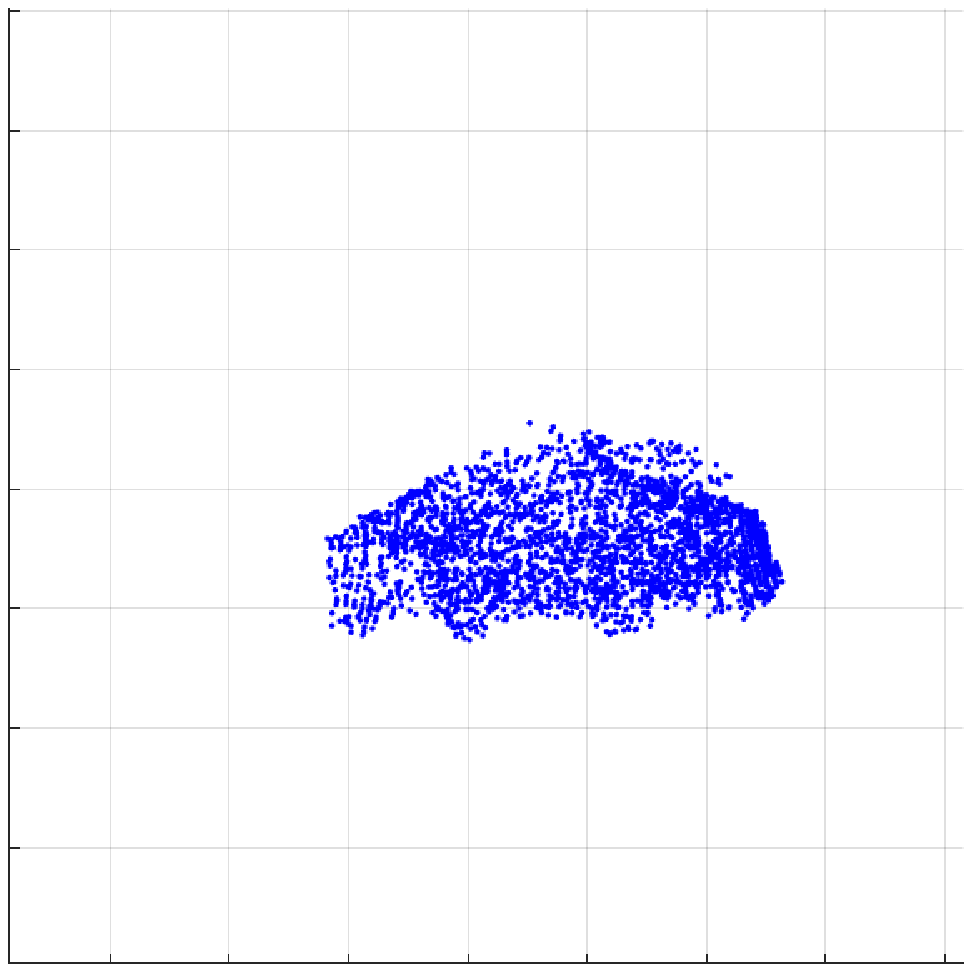}
			\setlength{\epsfxsize}{1.7in}
			\epsffile{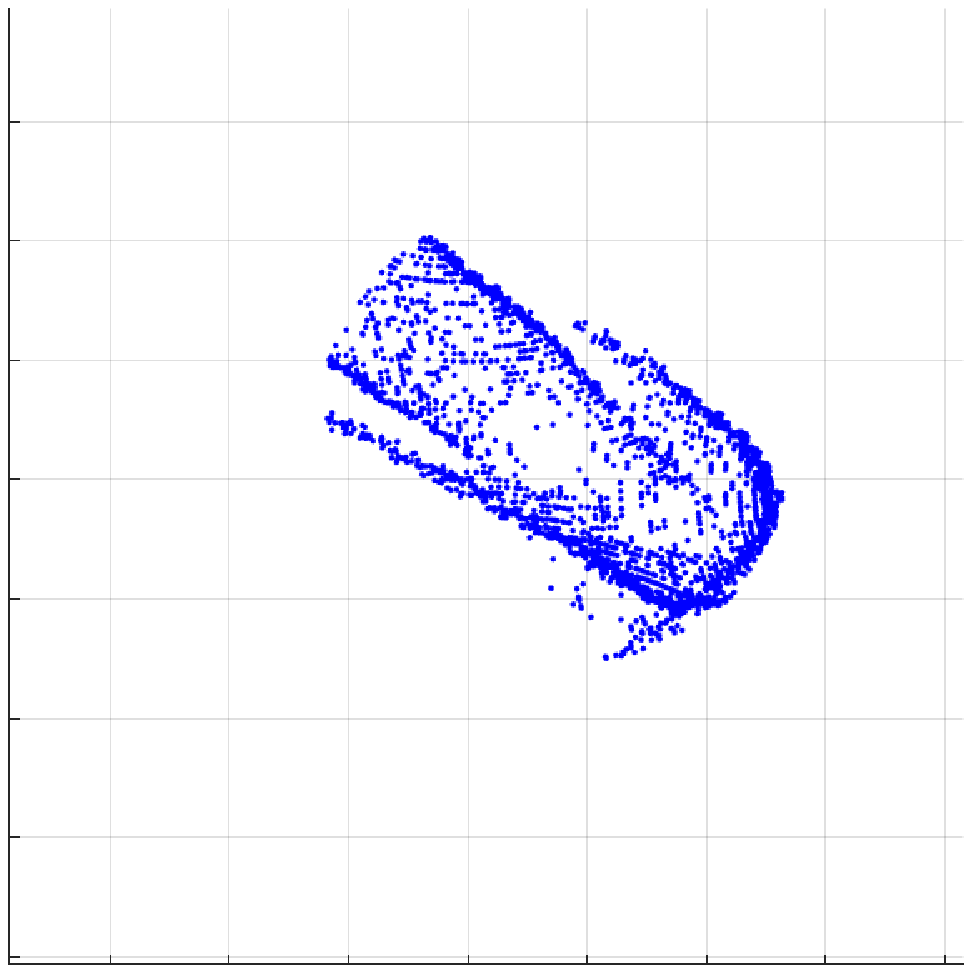}
		\end{center}
	\end{minipage}\label{fig_Lbox91_cam4_radar}\\
	
	\begin{minipage}{7.2in}
		\begin{center}
			\setlength{\epsfxsize}{1.7in}
			\epsffile{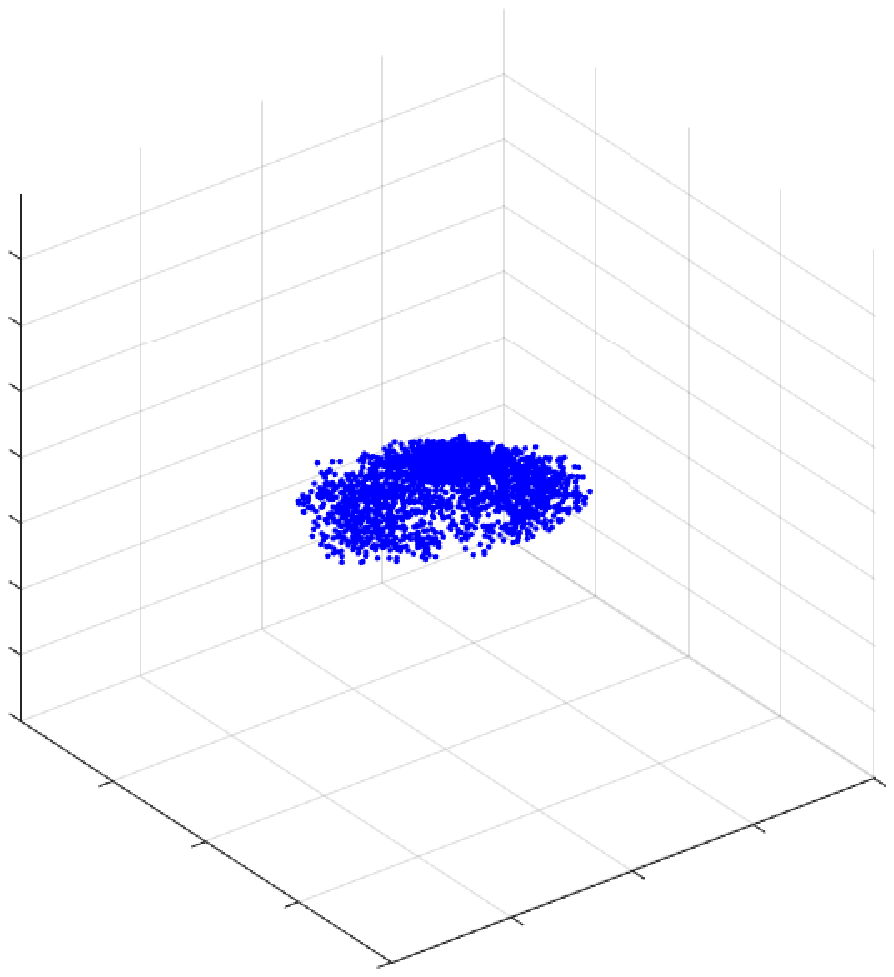}
			\setlength{\epsfxsize}{1.7in}
			\epsffile{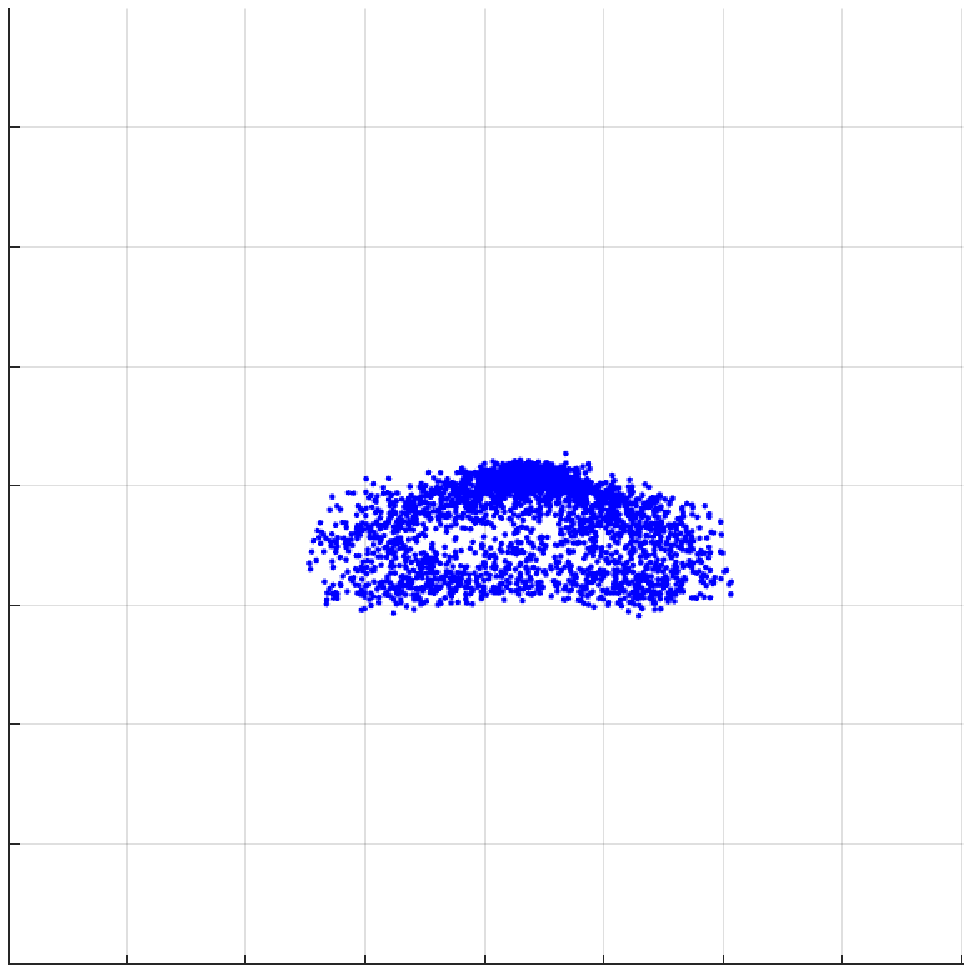}
			\setlength{\epsfxsize}{1.7in}
			\epsffile{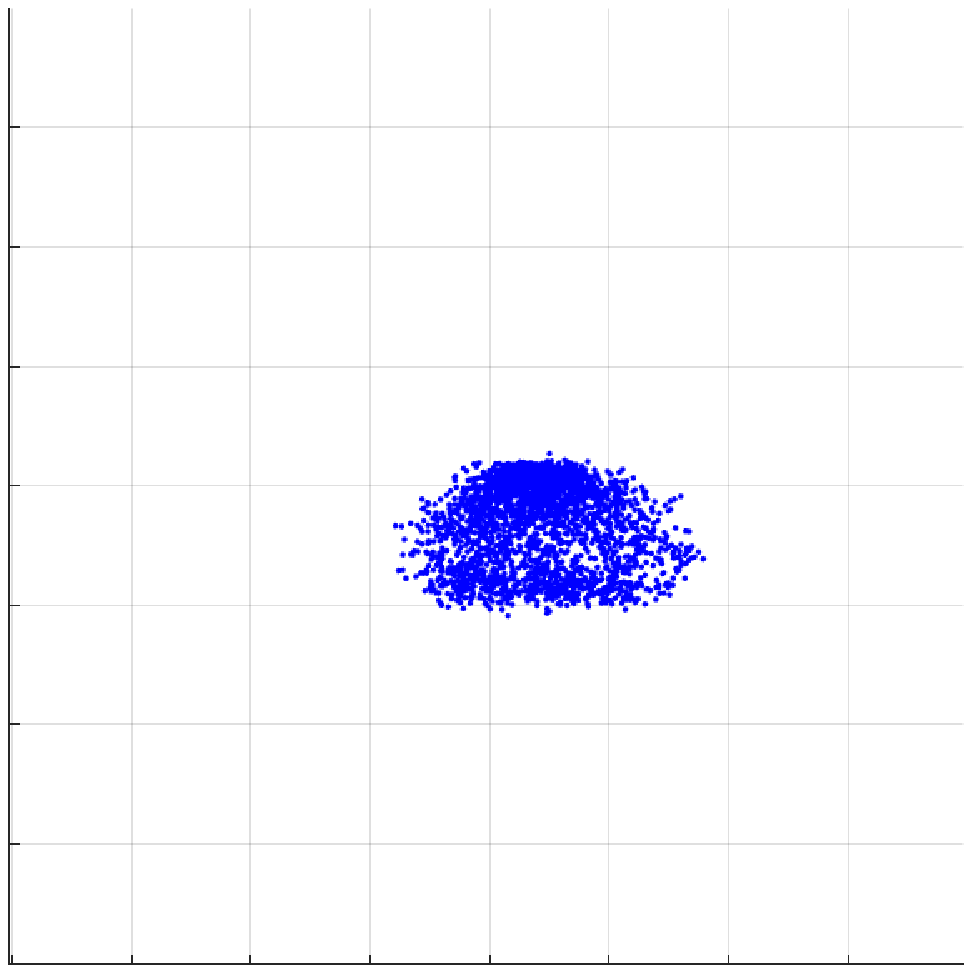}
			\setlength{\epsfxsize}{1.7in}
			\epsffile{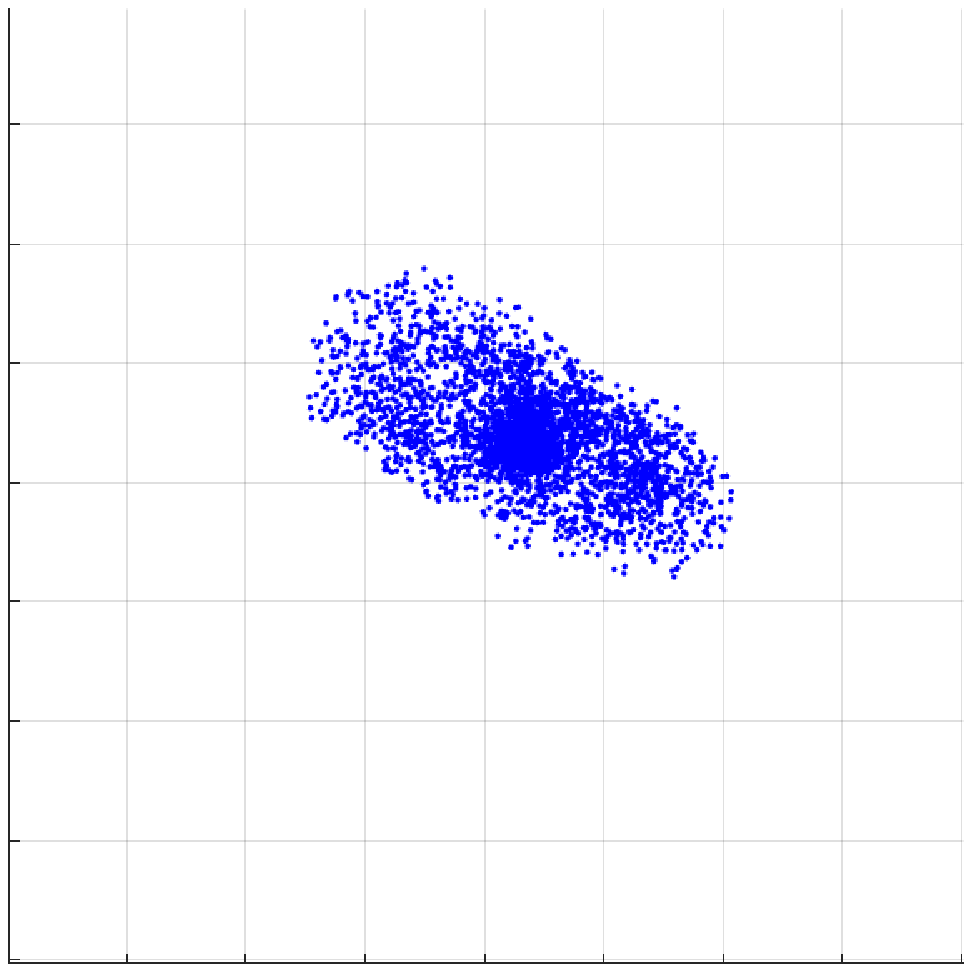}
		\end{center}
	\end{minipage}\vspace{0.05in}\label{fig_Lbox91_cam4_fake-depth}\\
	\begin{minipage}{7.2in}
		\begin{center}
			\setlength{\epsfxsize}{1.7in}
			\epsffile{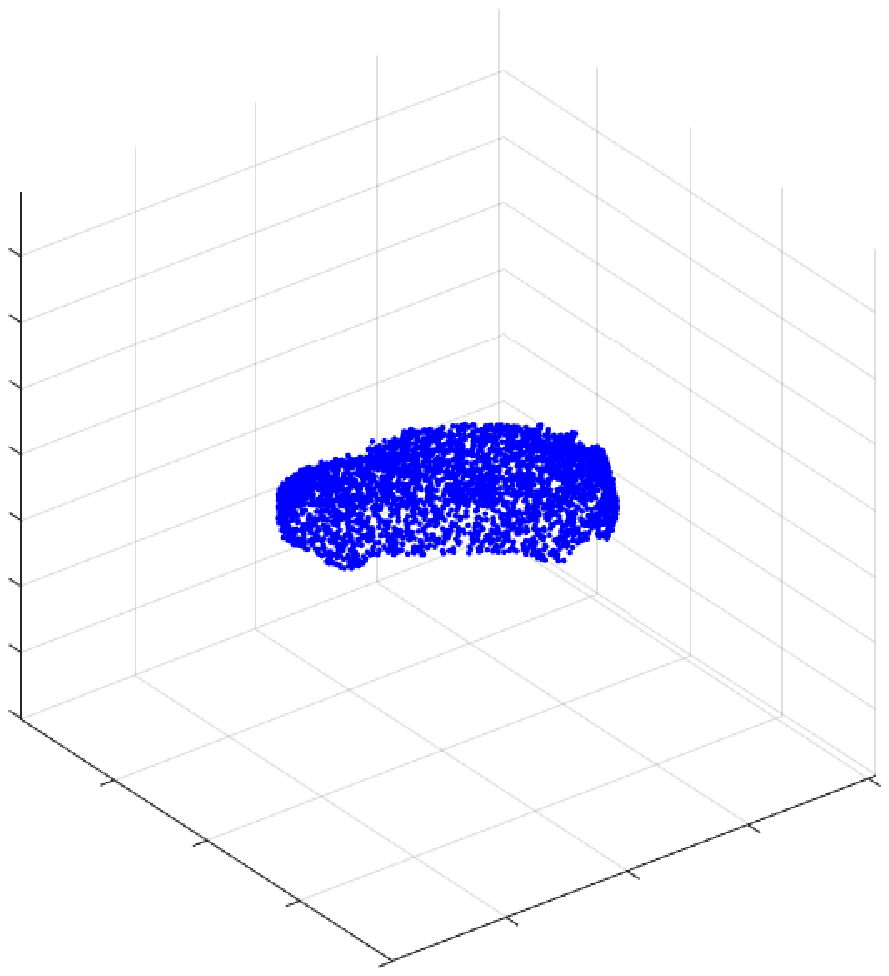}
			\setlength{\epsfxsize}{1.7in}
			\epsffile{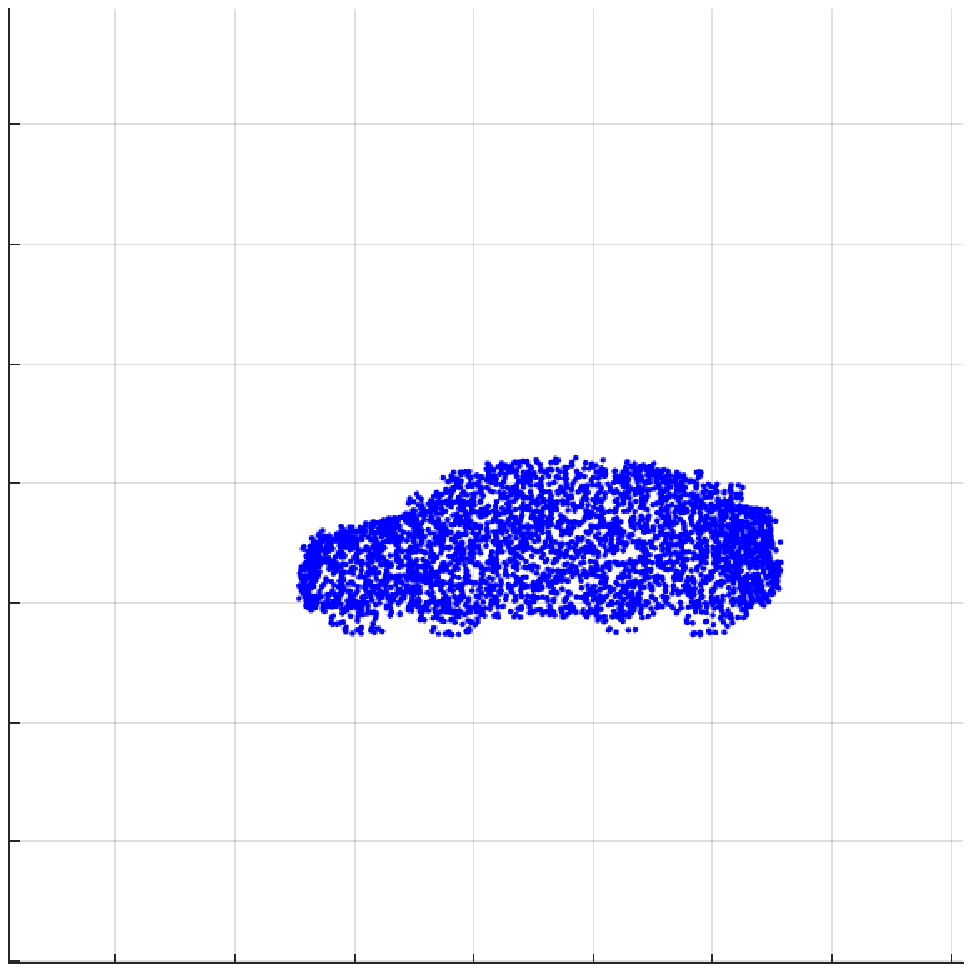}
			\setlength{\epsfxsize}{1.7in}
			\epsffile{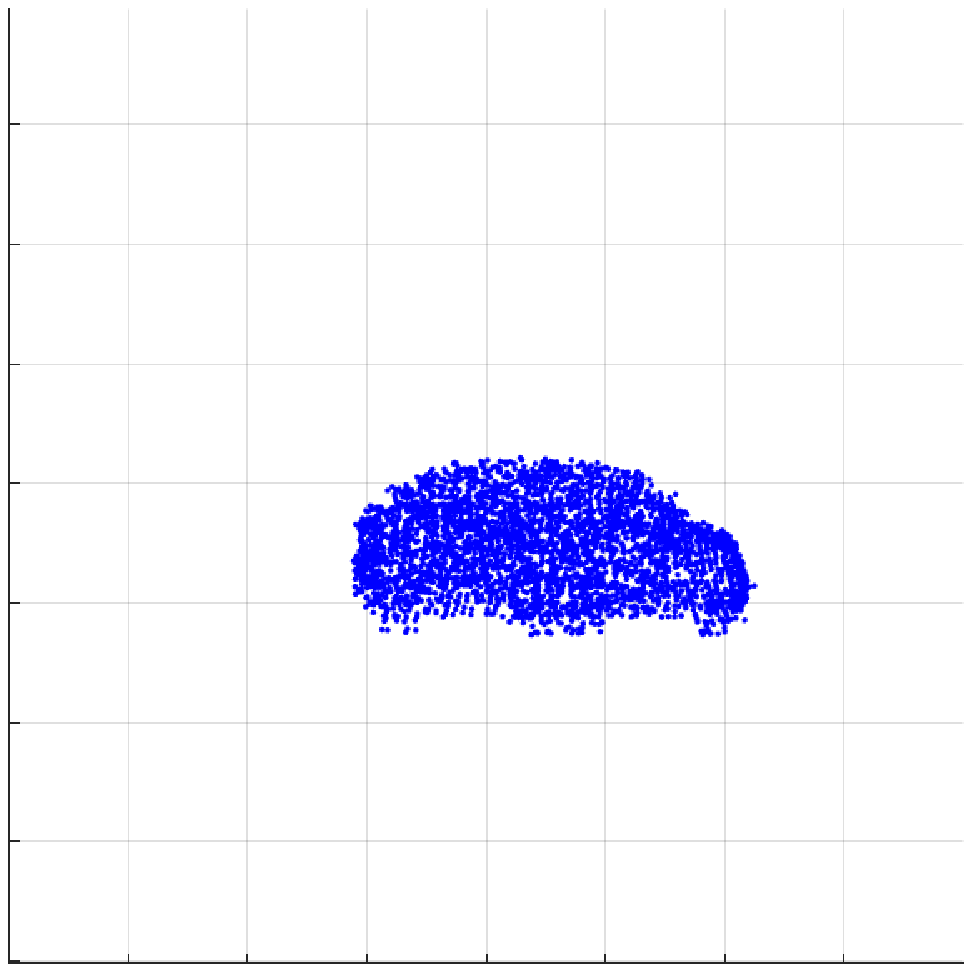}
			\setlength{\epsfxsize}{1.7in}
			\epsffile{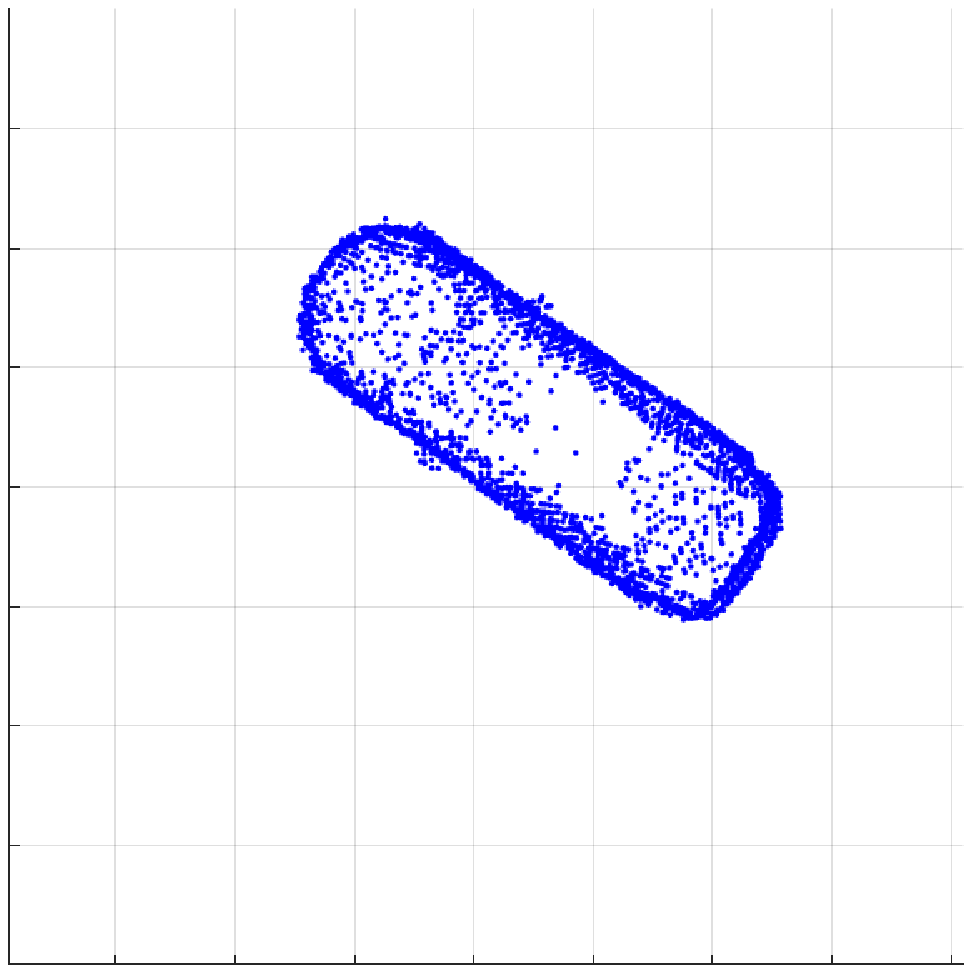}
		\end{center}
		\label{fig_Lbox91_cam4_real-depth}
	\end{minipage}
	%}
	\caption{3DRIMR Stage 2's point clouds in car experiments. The 1st row shows the input point clouds of a car from different viewpoints (i.e., inputs to the generator network in Stage 2). The 2nd row shows the output point clouds. The 3rd row shows the ground truth point clouds. Counting from the left, the 1st column lists point clouds shown in 3D space. The 2nd column lists the front views of point clouds. The 3rd column shows the left views of point clouds. The 4th column shows the top view of point clouds. }
	\label{fig_stage2_car_pc}
\end{figure*}

\subsection{Evaluation Results}

We compare 3DRIMR's results in Stage 1 against HawkEye  \cite{HawkEye} as HawkEye's goal is to  
generate 2D depth images only, 
and we compare 3DRIMR's final results against a double stacked PointNet architecture of PCN \cite{yuan2018pcn}, as PCN's goal 
is to generate 3D shapes in point cloud format based on sparse and partial point clouds.
Note that PCN does not rely on a GAN-based architecture but uses FoldingNet \cite{yang2017foldingnet}.
Example scenes of car and L-Box experiments are shown in Fig. \ref{fig_car_ex} and Fig. \ref{fig_Lbox_ex} respectively.
Four colored dots in each figure show four viewpoints of radar sensor or camera.

\subsubsection{Stage 1 results}

We show 3DRIMR's performance in car experiments 
in Fig. \ref{fig_stage1_car}.
%, in Lbox experiment test on the real data in 
%and Fig. \ref{fig_stage1_realLbox}.
%, and in Lbox experiment test on the synthesized data in Fig. \ref{fig_stage1_synLbox}. 
We see that visually 3DRIMR can accurately predict an object's shape, size, and orientation,
and it can also estimate the distances between an object's surface points and radar receiver.
% according to the consistency of depth intensity of predicted and ground truth depth images in all these 3 cases.
Although most of our training data is synthesized data, 3DRIMR can still predict depth images well based on the real data.
We observe similar performance from the experiments with L-Box. 

Note that we only use 2-snapshot 3D intensity data in our experiments whereas HawkEye  \cite{HawkEye} uses $64$ SAR snapshots along elevation, 
but our system's Stage 1 results are still comparable with those of HawkEye. 
To prove that, we evaluate the same metrics as those in \cite{HawkEye}.
%, and full detailed description of these metrics can be found in the supplementary document of paper  \cite{HawkEye}. 
Table \ref{stage1-res-table} shows the median errors of 3DRIMR's Stage 1 results compared against those in \cite{HawkEye}. 
Since the size of L-box is different from the size of cars, we scale it to the average size of cars when calculating length, width, and height errors.
%For the reason that we only have the synthesized data in our car experiment, and the real data in our Lbox experiment is collected only in the indoor environment,
%to compatiblely compare the performance between ours and HawkEye's, 
In the comparison, we calculate the average errors of HawkEye's three experiment settings (i.e., 
clean air, fog, and synthesized data).
Table \ref{stage1-res-table} shows that our system outperforms HawkEye in terms of range and orientation prediction.
Compared with HawkEye, our car predication's length error is $29\%$ larger, and our L-box prediction's 
length error is about $50\%$ smaller.
The errors in height and $\%$ Fictitious Reflections in both our car prediction and HawkEye's are very similar, but these errors in our L-box prediction are $67\%$ and $89\%$ smaller.
The errors in width and $\%$ Surface Missed of ours and HawkEye are similar. 
%in these 3 experiments have small differences.
In sum, 3DRIMR Stage 1's performance is comparable with that of HawkEye. Recall that Stage 1's results are just 3DRIMR's intermediate results. Next we evaluate 3DRIMR's final prediction results.

% table for stage1
\renewcommand{\arraystretch}{1.5} %
\begin{table*}[tp]	
	\centering
%	\fontsize{6.5}{8}\selectfont
	\fontsize{6}{8}\scriptsize
	\begin{tabular}{|c||c|c|c|c|c|c|c|}
		\hline
		Method&Error in Ranging&Error in Length&Error in Width&Error in Height&Error in Orientation&\% Fictitious Reflections&\% Surface Missed\cr
		\hline
		\hline
		3DRIMR-Cars           &16 cm     &84 cm               &37 cm        &10 cm             &  \bm{$4.8^{\circ} $}        &1.9 \%        &15.4 \% \cr\hline
		3DRIMR-Lbox-scaled &{\bf 8 cm}&{\bf 32 cm}       &{\bf 34 cm}&{\bf 3 cm }     &$12.4^{\circ} $                 &{\bf 0.2 \%}&16.1 \% \cr\hline
		HawkEye-avg      &34 cm     &65 cm                &37 cm       &9 cm              & $28.7^{\circ} $                &1.8 \%        &{\bf 12.8 \% }\cr\hline
	\end{tabular}
	\captionsetup{font={scriptsize}}
	\caption{Quantitative Results of 3DRIMR's Stage 1, compared with HawkEye \cite{HawkEye}.}
	\label{stage1-res-table}
\end{table*}

\subsubsection{Stage 2 results}

Fig. \ref{fig_stage2_car_pc} shows our final output point clouds compared with input point clouds and groud truth point clouds.
We can see that due to predict errors in Stage 1, the input point clouds to Stage 2, 
which are simply the unions of coarse point clouds from 4-view depth images, 
are discontinuous and have many incorrect points.
%That's a big challenge to reconstruct point clouds from point clouds with many fake points in them.
To the best of our knowledge, the state of art research only studies reconstructing point clouds from sparse and partial point clouds with missing points, but not dealing with systematic incorrect points which might give slanted or even wrong shapes. 
% at least all points of them are correct.
However, our system can perform reasonably well even based point clouds with incorrect points.
As shown in Fig. \ref{fig_stage2_car_pc}, the point clouds after 3DRIMR's Stage 2 are continuous and complete,
compared with the input point clouds.

We compare Stage 2 against two baseline methods.
\begin{itemize}
	\item \textbf{Double-PointNet Network (DPN)}.
	This is modified from our Stage 2 structure by removing $\bf{D_{p2p}}$, i.e., this is not a GAN architecture.
	This one is used to examine the effectiveness of GAN in training.
	
	\item \textbf{Chamfer-Loss-based Network (CLN)}. 
	This is the network modified from our Stage 2 structure by removing IoU loss from training process. 
	This one is used to examine whether IoU loss can help with the training. 
\end{itemize}

Table \ref{tab_stage2} shows the average and standard deviation of Chamfer Distances, IoUs, and F-Scores of 3DRIMR 
and the two baseline methods.
We can see that in both cars and L-box experiments, 3DRIMR always performs best among three methods in terms of all the evaluation metrics.
These results show that GAN architecture can indeed improve the generator's performance. 
In addition, Chamfer-Loss-based Network has the worst performance, which shows that our IoU loss can significantly improve the network's performance.

% stage2 figures end ===============================

% table for stage2
\renewcommand{\arraystretch}{1.5} 
\begin{table}[tp]	
	\centering
	\fontsize{6.5}{8}\scriptsize
	\begin{tabular}{|c|c|c|c|c|c|c|}
		\hline
		\multirow{2}{*}{Method}&
		\multicolumn{2}{c|}{Chamfer Distance}&\multicolumn{2}{c|}{ IoU}&\multicolumn{2}{c|}{ F-score}\cr\cline{2-7}
		&avg.&std.&avg.&std.&avg.&std.\cr
		\hline
		\hline
		3DRIMR-Cars              &{\bf 0.0789} &{\bf 0.0411}   &{\bf 0.0129}      &{\bf 0.0052} &{\bf 0.0841}    &{\bf 0.0322}    \cr\hline
		DPN-Cars  &0.1164         &0.1358           &0.0121            &0.0061            &0.0806           &0.0389          \cr\hline
		CLN-Cars     &0.1485         &0.1843           &0.0120           &0.0071            &0.809            &0.0447           \cr\hline
		\hline
		3DRIMR-Lbox            &{\bf 0.0205} &{\bf 0.0036}   &{\bf 0.0937}      &{\bf 0.0124} &{\bf 0.5575}    &{\bf 0.0961}    \cr\hline
		DPN-Lbox&0.0303         &0.0265          &0.0862             &0.0268         &0.4808           &0.1576             \cr\hline
		CLN-Lbox   &0.0367          &0.0558          &0.0845            &0.0281          &0.5103           &0.1774              \cr\hline
	\end{tabular}
	\captionsetup{font={scriptsize}}
	\caption{Quantitative Results of Stage 2, compared with two baseline methods.}
	\label{tab_stage2}
\end{table}

% --------------------------------------------------------------------------------------------------------------------------------------------------------------------------------

\section{Conclusions and Future work}\label{sec_conclusion}

We have proposed 3DRIMR, a deep learning architecture that reconstructs 3D object shapes in point cloud format based on raw mmWave radar
signals. 3DRIMR is a conditional GAN based architecture. This architecture takes advantage of 3D convolutional operation 
and point cloud's efficiency of representing 3D shapes. Our experiments have shown its effectiveness in reconstructing 3D objects based 
on two snapshots of a commodity mmWave sensor. For future work, we will further improve the design of the generator 
network of 3DRIMR's Stage 2 to improve the generated 3D shapes. We will also conduct large scale experiments to further improve
our design.

%%
%% REFERENCES
%%
\bibliographystyle{IEEEtran}
\bibliography{references}

\begin{thebibliography}{10}
\providecommand{\url}[1]{#1}
\csname url@samestyle\endcsname
\providecommand{\newblock}{\relax}
\providecommand{\bibinfo}[2]{#2}
\providecommand{\BIBentrySTDinterwordspacing}{\spaceskip=0pt\relax}
\providecommand{\BIBentryALTinterwordstretchfactor}{4}
\providecommand{\BIBentryALTinterwordspacing}{\spaceskip=\fontdimen2\font plus
\BIBentryALTinterwordstretchfactor\fontdimen3\font minus
  \fontdimen4\font\relax}
\providecommand{\BIBforeignlanguage}[2]{{%
\expandafter\ifx\csname l@#1\endcsname\relax
\typeout{** WARNING: IEEEtran.bst: No hyphenation pattern has been}%
\typeout{** loaded for the language `#1'. Using the pattern for}%
\typeout{** the default language instead.}%
\else
\language=\csname l@#1\endcsname
\fi
#2}}
\providecommand{\BIBdecl}{\relax}
\BIBdecl

\bibitem{HawkEye}
J.~Guan, S.~Madani, S.~Jog, S.~Gupta, and H.~Hassanieh, ``Through fog
  high-resolution imaging using millimeter wave radar,'' in \emph{IEEE CVPR
  2020}.

\bibitem{mobisys20smoke}
C.~X. Lu, S.~Rosa, P.~Zhao, B.~Wang, C.~Chen, J.~A. Stankovic, N.~Trigoni, and
  A.~Markham, ``See through smoke: robust indoor mapping with low-cost mmwave
  radar,'' in \emph{ACM MobiSys 2020}.

\bibitem{superrf}
S.~Fang and S.~Nirjon, ``Superrf: Enhanced 3d rf representation using
  stationary low-cost mmwave radar,'' in \emph{Proc. of 2020 Intl Conf on
  Embedded Wireless Systems and Networks}.

\bibitem{iwr6843}
Texas-Instruments. Iwr6843isk, 2021. \url {https://www.ti.com/tool/IWR6843ISK}.

\bibitem{vandersmissen2018indoor}
B.~Vandersmissen, N.~Knudde, A.~Jalalvand, I.~Couckuyt, A.~Bourdoux,
  W.~De~Neve, and T.~Dhaene, ``Indoor person identification using a low-power
  fmcw radar,'' \emph{IEEE Transactions on Geoscience and Remote Sensing},
  vol.~56, no.~7, pp. 3941--3952, 2018.

\bibitem{yang2020mu}
X.~Yang, J.~Liu, Y.~Chen, X.~Guo, and Y.~Xie, ``Mu-id: Multi-user
  identification through gaits using millimeter wave radios,'' in \emph{IEEE
  INFOCOM 2020}.

\bibitem{mamandipoor201460}
B.~Mamandipoor, G.~Malysa, A.~Arbabian, U.~Madhow, and K.~Noujeim, ``60 ghz
  synthetic aperture radar for short-range imaging: Theory and experiments,''
  in \emph{2014 48th Asilomar Conference on Signals, Systems and
  Computers}.\hskip 1em plus 0.5em minus 0.4em\relax IEEE, 2014, pp. 553--558.

\bibitem{national2018airport}
E.~National Academies~of Sciences, Medicine \emph{et~al.}, \emph{Airport
  Passenger Screening Using Millimeter Wave Machines: Compliance with
  Guidelines}.\hskip 1em plus 0.5em minus 0.4em\relax National Academies Press,
  2018.

\bibitem{ghasr2016wideband}
M.~T. Ghasr, M.~J. Horst, M.~R. Dvorsky, and R.~Zoughi, ``Wideband microwave
  camera for real-time 3-d imaging,'' \emph{IEEE Transactions on Antennas and
  Propagation}, vol.~65, no.~1, pp. 258--268, 2016.

\bibitem{sheen2007near}
D.~M. Sheen, D.~L. McMakin, and T.~E. Hall, ``Near field imaging at microwave
  and millimeter wave frequencies,'' in \emph{2007 IEEE/MTT-S International
  Microwave Symposium}.\hskip 1em plus 0.5em minus 0.4em\relax IEEE, 2007, pp.
  1693--1696.

\bibitem{yuan2018pcn}
W.~Yuan, T.~Khot, D.~Held, C.~Mertz, and M.~Hebert, ``Pcn: Point completion
  network,'' in \emph{2018 International Conference on 3D Vision (3DV)}.\hskip
  1em plus 0.5em minus 0.4em\relax IEEE, 2018, pp. 728--737.

\bibitem{qi2016pointnet}
C.~R. Qi, H.~Su, K.~Mo, and L.~J. Guibas, ``Pointnet: Deep learning on point
  sets for 3d classification and segmentation,'' \emph{arXiv preprint
  arXiv:1612.00593}, 2016.

\bibitem{sun2020lidaus}
Y.~Sun, D.~Xu, Z.~Huang, H.~Zhang, and X.~Liang, ``Lidaus: Localization of iot
  device via anchor uav slam,'' in \emph{IEEE IPCCC 2020}.

\bibitem{yang20173d}
B.~Yang, H.~Wen, S.~Wang, R.~Clark, A.~Markham, and N.~Trigoni, ``3d object
  reconstruction from a single depth view with adversarial learning,'' in
  \emph{IEEE International Conference on Computer Vision Workshops}, 2017.

\bibitem{dai2017shape}
A.~Dai, C.~Ruizhongtai~Qi, and M.~Nie{\ss}ner, ``Shape completion using
  3d-encoder-predictor cnns and shape synthesis,'' in \emph{IEEE CVPR 2017}.

\bibitem{sharma2016vconv}
A.~Sharma, O.~Grau, and M.~Fritz, ``Vconv-dae: Deep volumetric shape learning
  without object labels,'' in \emph{European Conference on Computer
  Vision}.\hskip 1em plus 0.5em minus 0.4em\relax Springer, 2016, pp. 236--250.

\bibitem{smith2017improved}
E.~J. Smith and D.~Meger, ``Improved adversarial systems for 3d object
  generation and reconstruction,'' in \emph{Conference on Robot
  Learning}.\hskip 1em plus 0.5em minus 0.4em\relax PMLR, 2017, pp. 87--96.

\bibitem{timmwave}
Texas-Instruments. Introduction to mmwave radar sensing: Fmcw radars. \url
  {https://training.ti.com/node/1139153}.

\bibitem{fan2017point}
H.~Fan, H.~Su, and L.~J. Guibas, ``A point set generation network for 3d object
  reconstruction from a single image,'' in \emph{IEEE CVPR 2017}.

\bibitem{qi2017pointnet++}
C.~R. Qi, L.~Yi, H.~Su, and L.~J. Guibas, ``Pointnet++: Deep hierarchical
  feature learning on point sets in a metric space,'' \emph{arXiv preprint
  arXiv:1706.02413}, 2017.

\bibitem{kar2017learning}
A.~Kar, C.~H{\"a}ne, and J.~Malik, ``Learning a multi-view stereo machine,'' in
  \emph{Proceedings of the 31st International Conference on Neural Information
  Processing Systems}, 2017, pp. 364--375.

\bibitem{paschalidou2018raynet}
D.~Paschalidou, O.~Ulusoy, C.~Schmitt, L.~Van~Gool, and A.~Geiger, ``Raynet:
  Learning volumetric 3d reconstruction with ray potentials,'' in \emph{IEEE
  CVPR 2018}.

\bibitem{ji2017surfacenet}
M.~Ji, J.~Gall, H.~Zheng, Y.~Liu, and L.~Fang, ``Surfacenet: An end-to-end 3d
  neural network for multiview stereopsis,'' in \emph{Proceedings of the IEEE
  International Conference on Computer Vision}, 2017, pp. 2307--2315.

\bibitem{wu2016learning}
J.~Wu, C.~Zhang, T.~Xue, W.~T. Freeman, and J.~B. Tenenbaum, ``Learning a
  probabilistic latent space of object shapes via 3d generative-adversarial
  modeling,'' \emph{arXiv preprint arXiv:1610.07584}, 2016.

\bibitem{kong2017using}
C.~Kong, C.-H. Lin, and S.~Lucey, ``Using locally corresponding cad models for
  dense 3d reconstructions from a single image,'' in \emph{IEEE CVPR 2017}.

\bibitem{wang2018pixel2mesh}
N.~Wang, Y.~Zhang, Z.~Li, Y.~Fu, W.~Liu, and Y.-G. Jiang, ``Pixel2mesh:
  Generating 3d mesh models from single rgb images,'' in \emph{Proceedings of
  the European Conference on Computer Vision (ECCV)}, 2018, pp. 52--67.

\bibitem{encoder-decoder}
V.~Badrinarayanan, A.~Kendall, and R.~Cipolla, ``Segnet: A deep convolutional
  encoder-decoder architecture for image segmentation,'' \emph{IEEE
  Transactions on Pattern Analysis and Machine Intelligence}, vol.~39, no.~12,
  pp. 2481--2495, 2017.

\bibitem{VGG}
K.~Simonyan and A.~Zisserman, ``Very deep convolutional networks for
  large-scale image recognition,'' 2015.

\bibitem{dca1000evm}
Texas-Instruments. Dca1000evm. \url {https://www.ti.com/tool/DCA1000EVM}.

\bibitem{fidler20123d}
S.~Fidler, S.~Dickinson, and R.~Urtasun, ``3d object detection and viewpoint
  estimation with a deformable 3d cuboid model,'' \emph{Advances in neural
  information processing systems}, 2012.

\bibitem{zed}
STEREOLABS. Zed mini datasheet 2019 rev1. \url
  {https://cdn.stereolabs.com/assets/datasheets/zed-mini-camera-datasheet.pdf}.

\bibitem{yang2017foldingnet}
Y.~Yang, C.~Feng, Y.~Shen, and D.~Tian, ``Foldingnet: Interpretable
  unsupervised learning on 3d point clouds,'' \emph{arXiv preprint
  arXiv:1712.07262}, vol.~2, no.~3, p.~5, 2017.

\end{thebibliography}

\end{document}